\documentclass[useAMS,usenatbib]{mn2e}

\usepackage{graphicx}
\usepackage{txfonts}
\usepackage{natbib}
\usepackage{color}
\usepackage{longtable,lscape}

\def\etal{\mbox{\rm et al.\ }}

\def\teff{\mbox{T$_{\rm eff}$}}
\def\logg{\mbox{log~{\it g}}}
\def\vmicro{\mbox{$\xi_{\rm t}$}}
\def\kmsec{\mbox{km~s$^{\rm -1}$}}

\def\vsini{\mbox{$v~{\rm sin}{\it i}$}}

\newcommand\aj{AJ}
\newcommand\apj{ApJ}
\newcommand\apjs{ApJS}
\newcommand\apss{Ap\&SS}
\newcommand\aap{A\&A}
\newcommand\mnras{MNRAS}
\newcommand\apjl{ApJ}

\newcommand\nat{Nature}
\newcommand\aaps{A\&AS}

\title[]{
Helium enhanced stars and multiple populations along the horizontal branch of NGC\,2808: direct spectroscopic measurements\thanks{Based on data collected at the European Southern Observatory with the FLAMES/GIRAFFE and FLAMES/UVES spectrographs, under the program 086.D-0141.}}
\author[A.\, F.\, Marino et al.]
{A.\, F.\, Marino$^{1}$\thanks{E-mail:amarino@mso.anu.edu.au},
A.\, P.\, Milone$^{1}$,
N.\, Przybilla$^{2,3}$,
M.\, Bergemann$^{4}$,
K.\, Lind$^{5}$,
\newauthor
M.\, Asplund$^{1}$,
S.\, Cassisi$^{6}$,
M.\, Catelan$^{7}$,
L.\, Casagrande$^{1}$,
A.\,A.\,R.\, Valcarce$^{8}$,
L.\, R.\, Bedin$^{9}$,
\newauthor
C.\, Cort\'{e}s$^{10}$,
F.\, D'Antona$^{11}$,
H.\, Jerjen$^{1}$,
G.\, Piotto$^{9,12}$,
K.\, Schlesinger$^{1}$,
M.\, Zoccali$^{7}$,
\newauthor
R.\, Angeloni$^{7,13,14}$
\\
$^{1}$Research School of Astronomy \& Astrophysics, Australian National
University, Mt Stromlo Observatory, via Cotter Rd, Weston, ACT 2611, Australia \\
$^{2}$Institute for Astro- and Particle Physics, University of Innsbruck,
Technikerstr. 25/8, 6020, Innsbruck, Austria\\
$^{3}$Dr. Karl Remeis-Observatory \& ECAP, University Erlangen-N\"urnberg, Sternwartstr. 7,
96049, Bamberg, Germany\\
$^{4}$Max-Planck-Institut f\"ur Astrophysik Karl-Schwarzschild-Str.\, 1 85741 Garching bei M\"unchen Germany \\ 
$^{5}$Institute of Astronomy, University of Cambridge, Madingley Road, Cambridge, CB3 0HA, UK \\
$^{6}$INAF-Osservatorio Astronomico di Teramo, via M.\, Maggini, 64100 Teramo, Italy \\
$^{7}$Instituto de Astrofis{\'{i}}ca, Pontificia Universidad Cat{\'{o}}lica de Chile, Av. Vicu\~{n}a Mackenna 4860, 782-0436 Macul, Santiago, Chile\\
$^{8}$Universidade Federal do Rio Grande do Norte, Departamento de F\'{i}sica, 59072-970, Natal, RN, Brazil\\
$^{9}$INAF-Osservatorio Astronomico di Padova, Vicolo dell'Osservatorio 5, Padova I-35122, Italy\\
$^{10}$Universidad Metropolitana de Ciencias de la Edicaci\'{o}n, Santiago, Chile\\
$^{11}$INAF-Osservatorio Astronomico di Roma, via Frascati 33, I-00040 Monteporzio, Italy\\
$^{12}$Dipartimento di Fisica e Astronomia `Galileo Galilei', Universit\`a di Padova, Vicolo dell'Osservatorio 3, Padova, I-35122, Padova, Italy.\\
$^{13}$Department of Electrical Engineering, Center for Astro-Engineering, Pontificia Universidad Cat\'{o}lica de Chile, Av. Vicu\~{n}a Mackenna 4860, 782-0436 Macul, Santiago, Chile\\
$^{14}$The Milky Way Millennium Nucleus, Av. Vicu\~{n}a Mackenna 4860, 782-0436 Macul, Santiago, Chile\\
}

\begin{document}


\pagerange{\pageref{firstpage}--\pageref{lastpage}} \pubyear{2013}

\maketitle

\label{firstpage}

\begin{abstract}  
We present an abundance analysis of 96 horizontal branch (HB) stars in NGC\,2808, a globular cluster exhibiting a complex multiple stellar population pattern. These stars are distributed in different portions of the HB and cover a wide range of temperature. By studying the chemical abundances of this sample, we explore the connection between HB morphology and the chemical enrichment history of multiple stellar populations. For stars lying on the red HB, we use GIRAFFE and UVES spectra to determine Na, Mg, Si, Ca, Sc, Ti, Cr, Mn, Fe, Ni, Zn, Y, Ba, and Nd abundances. 
For colder, blue HB stars, we derive abundances for Na, primarily from GIRAFFE spectra. We were also able to measure direct NLTE He abundances for a subset of these blue HB stars with temperature higher than $\sim$9000~K. Our results show that: {\it (i)} HB stars in NGC\,2808 show different content in Na depending on their position in the color-magnitude diagram, with blue HB stars having higher Na than red HB stars; {\it (ii)} the red HB is not consistent with an uniform chemical abundance, with slightly warmer stars exhibiting a statistically significant higher Na content; and {\it (iii)} our subsample of blue HB stars with He abundances shows evidence of enhancement with respect to the predicted primordial He content by $\Delta Y=+0.09 \pm 0.01$. Our results strongly support theoretical models that predict He enhancement among second generation(s) stars in globular clusters and provide observational constraints on the second-parameter governing HB morphology.
\end{abstract}


\section{Introduction}
\label{introduction}

Globular clusters (GCs) exhibit a range of horizontal branch (HB) morphology, with some clusters showing red, blue, extended blue, or even multimodal HBs. While variations in metallicity are proposed as the first parameter governing the HB, these alone cannot explain the wide range of observed structure. Many GC properties, such as age, mass loss, and He content, have been proposed to be the second-parameter, but none has been fully successful (see Catelan 2009 for a review).

Observations of multiple stellar populations in GCs allow us to look at the second-parameter problem from a new perspective. Recent studies showed that the color-magnitude diagram (CMD) of GCs, from the main sequence (MS) up to the red-giant branch (RGB), if analysed in appropriate filters, are composed of different sequences, which correspond to different stellar populations with a range of light-element and (possibly) He abundances. Furthermore, the multiple HB components revealed in the CMD of several GCs have been tentatively assigned to different stellar generations by many authors (e.g.\, D'Antona et al.\ 2004, 2005; Piotto et al.\ 2007; Busso et al.\ 2007; Milone et al.\ 2008; Milone 2013). 

The association of chemical inhomogeneity and HB structure was originally advanced by Norris (1981) and Smith \& Norris (1983) in the context of CN variations, and by Catelan \& de Freitas Pacheco (1995) in the context of extra-mixed O-poor stars. He-enhanced stars are expected to populate the blue extreme of the HB (e.g.\, D'Antona et al.\ 2002), and to be O-depleted and Na-enhanced, as observed in the surface abundances of second-generation(s) GC stars. In some cases, the connection between chemistry and HB morphology is supported by the agreement between the fraction of stars on individual GC CMD sequences, associated with different chemical content, and the fraction in different HB segments (e.g.\ D'Antona et al.\ 2005; Piotto et al.\ 2007; Milone et al. 2008; Milone et al.\ 2012a).

A clear and direct confirmation of the chemistry/HB-morphology connection in GCs was presented in recent work on M\,4 (Marino et al.\ 2011a). Spectroscopic studies of the RGB revealed that two groups of stars are present in this cluster, one Na/CN-band strong (O-poor) and the other Na/CN-band weak (O-rich). These stellar populations define two different RGB sequences in the $B$ vs.\ $(U-B)$ CMD (see Marino et al.\ 2008). Analogous to the RGB stellar distribution, the HB is also bimodal, i.e.\, is well populated on both sides of the instability strip. Stars located on the blue side of the RR-Lyrae gap are Na-rich, and stars on the red side are Na-poor (see Fig.1a from Marino et al.\ 2011a). Successively, the connection between stellar populations and HB morphology in GCs  has been confirmed for NGC\,2808 (Gratton et al.\ 2011), NGC\,1851 (Gratton et al.\ 2012), 47\,Tucanae (Milone et al.\ 2012a, Gratton et al.\ 2013), NGC\,6397 (Lovisi et al.\ 2012), M\,5 (Gratton et al.\ 2013), and M\,22 (Marino et al. 2013).

The interplay of He-enhanced stars in GCs seems to be the only viable way to explain multiple MSs in GCs. In some cases, like $\omega$~Centauri and NGC\,2808, the He enrichment in second stellar generations is expected to be large (up to $Y$$\sim$0.40; Bedin et al.\ 2004; Norris 2004; Piotto et al. 2005; 2007; King et al.\ 2012), 
in other cases, like NGC\,6752 and NGC\,6397, the minimal separation in color between the sequences is consistent with much smaller He enhancements (Milone et al. 2013; 2012b). 

In this observational and theoretical framework, strong spectroscopic evidence of He-enhanced stars in a GC will both corroborate the proposed generational scenario to explain multiple MSs, and shed light on the second-parameter problem. Unfortunately, while abundance variations in $p$-capture elements, such as O and Na, have been widely observed, direct spectroscopic evidence of He enhancement in GCs is scarce, as these measurements are very difficult. Reliable He abundances can be measured for only a small fraction of HB stars with surface temperatures between $\sim$8000 and $\sim$11500~K. Hotter stars experience He settling, which results in abundances in the stellar atmosphere that are not representative of the initial He content (e.g.\ Behr 2003; Moehler et al.\ 2004; Fabbian et al.\ 2005). Cool stars do not typically have strong enough He lines in the optical. When they do, these He lines are chromospheric rather than photospheric; to determine a reliable abundance from these features requires complex models that take into account the chromospheric activity (Dupree et al. 2011). 

Pasquini et al. (2011) and Dupree et al. (2011) used near-infrared He lines in RGB stars of NGC\,2808 and $\omega$~Centauri, respectively, and found deeper and/or detectable lines for Na-rich stars. These results strongly suggest a higher He abundance for the Na-rich stars hosted in these two clusters. Pasquini et al. (2011) suggested a difference $\Delta(Y)\gtrsim$0.17 in He for the two RGBs in NGC\,2808, although they cannot provide the absolute abundances. A similar difference has also been estimated for two RGBs in $\omega$~Centauri, with absolute He abundances inferred from one near-infrared line analysed using appropriate chromospheric models (Dupree \& Avrett 2013).

Villanova et al. (2009) presented the first direct He abundance determinations from spectral lines not significantly affected by chromospheric activity for four blue HB stars in NGC\,6752 and six in M\,4. For three of the four stars in NGC\,6752, Na, O, and He were compatible with first-generation GC stars (primordial $Y$ and Na-poor/O-rich). For M\,4 the He content appeared enhanced with respect to the primordial value, by $\Delta(Y) \sim$0.04. 
Note however that these two clusters are not ideal to detect He enhancements. NGC\,6752 shows only a blue HB (HB ratio HBR=1.0, Harris 1996, updated as in 2010). This suggests that the different stellar populations observed on the RGB and MS (including the He-primordial first generation) are expected to distribute along the blue HB. In fact, first-generation stars have been found to populate the coldest segment of the blue HB of NGC\,6752 (Villanova et al. 2009). Furthermore, according to the MS splits observed by Milone et al. (2013), second-generation stars in NGC\,6752 and M\,4 will be enhanced in helium by only small amounts, which would be very difficult to detect spectroscopically.

The GC NGC\,2808 represents the ideal target to detect the presence of HB He enhanced stars in GCs. While its uniformity in metallicity suggests that this GC could be classified among {\it normal}\footnote{Based on Marino et al.\ (2012a), we consider GCs exhibiting internal variations in iron and heavy elements to be anomalous. Examples of these GCs are M22 (Marino et al. 2009, 2011b), NGC\,1851 (Yong \& Grundahl 2008; Carretta et al. 2011), and the more extreme $\omega$~Centauri (e.g.\, Norris \& Da Costa 1995; Johnson \& Pilachowski 2010; Marino et al.\ 2011c, 2012b).} GCs, i.e.\, those showing multiple stellar populations by means of variations in the chemical content of light elements (e.g. C, N, O, Na, Mg, Al) alone, photometric studies show that it is more complex. 
{\it Hubble Space Telescope} ({\it HST}) photometry has discovered a spectacular triple main sequence (MS), unambiguously revealing that NGC\,2808 hosts at least three different stellar generations (Piotto et al. 2007; Milone et al. 2012c,d). The RGB also exhibits evidence of non-singular populations, with large color spreads (Yong et al.\ 2008; Lee et al.\ 2009; Monelli et al.\ 2013). NGC\,2808 exhibits complex HB morphology, with multiple components and stars distributed along an extended blue tail (Sosin et al. 1997; Bedin et al. 2000, hereafter B00; Iannicola et al.\ 2009). Helium enhanced stars are expected to lie on the blue side of the instability strip, including the temperature range suitable for He measurements from spectral lines, and theoretical models require a large He enhancement, up to $Y$$\sim$0.38, to reproduce the CMD of this cluster (D'Antona et al. 2005; Piotto et al. 2007; Milone et al. 2012d). On the spectroscopic side, NGC\,2808 shows an extended Na-O anticorrelation (Carretta et al. 2006). In addition, Bragaglia et al. (2010) analysed one red and one blue star on the MS, finding that the blue was enhanced in N and Al, while being depleted in C. This chemical information suggests that the blue MS stars in NGC\,2808 may be helium enhanced. 

In this paper we analyse chemical abundances, including He, along the HB of NGC\,2808. The organisation of this paper is as follows: in Sect.~\ref{sec:data} we describe our sample and observations; Sect.~\ref{atm} explains our atmospheric parameter determination; Sect.~\ref{abunds} summarises our derived chemical abundances, with detailed results for Na and He described in Sects.~\ref{results} and \ref{heBHB}, respectively. We then discuss our results and repercussions for the second-parameter problem, in Sect.~\ref{discussion}. A summary of our results is presented in Sect.~\ref{conclusions}.

\section{Data}
\label{sec:data}

\subsection{The photometric catalogs}
\label{sect_phot}
To identify our stellar sample, we use the photometric catalogue of Momany et al.\ (2004), which has been obtained from $U$, $B$, and $V$ images collected with the Wide-Field Imager (WFI) mounted at the 2.2m ESO-MPI telescope at La Silla observatory, Chile. This catalogue has been used in previous spectroscopic studies of the HB of NGC\,2808 (Pace et al.\ 2006, Gratton et al.\ 2011). To estimate systematic errors related to the adopted photometric catalog, we also use photometry from B00, which was obtained from $U$, $B$, and $V$ images taken with the Danish Faint Object Spectrograph and Camera (DFOSC) mounted at the 1.54 ESO-Danish telescope at La Silla. 
 
The average reddening of NGC\,2808 is $E(B-V)=0.19$ (B00) and is not uniform across the field of view analysed in this paper. Significant star-to-star reddening variations have been observed in even a small $\sim$3\arcmin$\times$3\arcmin\ field around the cluster center (Milone et al.\ 2012c,d, Sect.~2).  

   \begin{figure*}
   \centering
   \includegraphics[width=17cm]{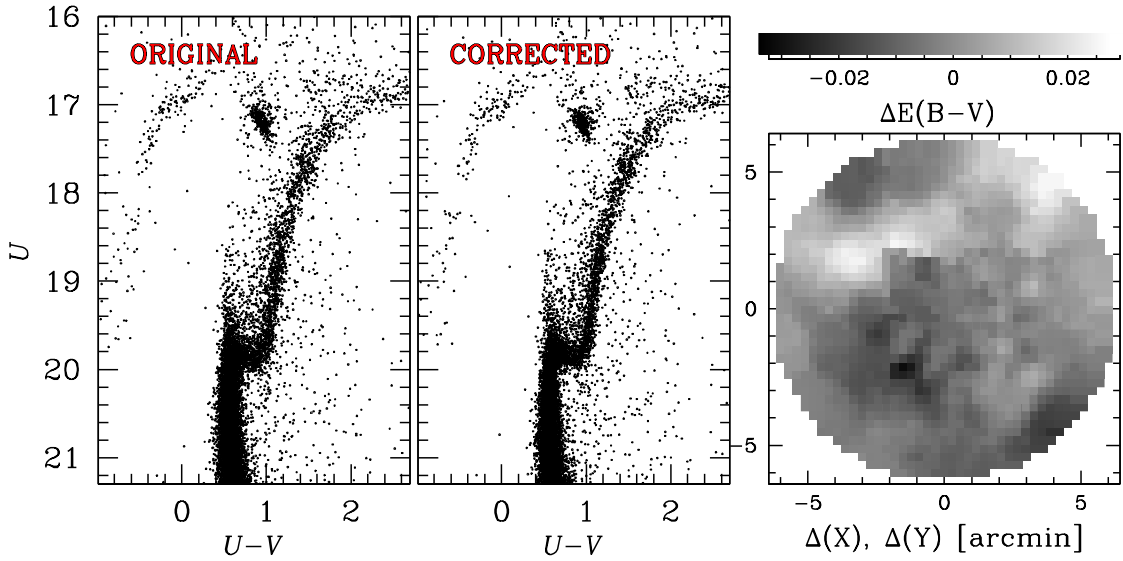}
\caption{{\it Left panels}: Comparison between the original 
$U$ versus $(U-V)$ CMD of NGC\,2808 (\textit{left-hand}) 
and the CMD corrected for differential reddening (\textit{right-hand}).
{\it Right Panel}: Map of the differential reddening and zero points effects in the 
11.25\arcmin$\times$11.25\arcmin\ field of view centered on NGC\, 2808.  
The gray scale reflects the magnitude of the variation in local
  reddening. }
         \label{fig:comp}
   \end{figure*}

A visual inspection of the CMD shown in the left panel of Fig.~\ref{fig:comp} reveals that the sequences, from MS to RGB, are spread by 
the interplay of several effects such as differential reddening, zero point variations, and sky concentration\footnote{Sky concentration is a spurious re-distribution of light in the focal plane due to the reflections of light at discontinuities in the optics (Manfroid et al. 2001). The WFI@2.2m camera is significantly affected by this and by zero point variations, as discussed in Bellini et al. (2009).}. To correct for these effects, we used the procedure described by Milone et al. (2012e). Briefly, we first draw a main-sequence ridge line by putting a spline through the median colors found in successive short intervals of magnitude, and iterating with sigma clipping. For each star, we then estimate how the observed stars in its vicinity may systematically lie to the red or the blue of the fiducial sequence; this systematic color and magnitude offset, measured along the reddening line, is indicative of the local differential reddening. This procedure treats the CMD as a single sequence. In the case of CMDs characterised by multiple sequences, like in NGC\,2808, this approximation results in larger internal uncertainties, but it does not systematically affect the correction (see Milone et al. 2012e). The final CMD, corrected for differential reddening and zero points is shown in the middle panel of Fig.~\ref{fig:comp}. Our corrections significantly decrease the internal photometric uncertainty, as shown by the comparison between the original and corrected CMDs of Fig.~\ref{fig:comp}. To give an idea of the size of our corrections, we show the spatial reddening variations in the field of NGC\,2808 in the right panel of Fig.~\ref{fig:comp}. We divide the whole field of view into 64$\times$64 boxes of 11.25\arcsec$\times$11.25\arcsec\ and calculate the average $\Delta~E(B-V)$ within each of them. In the resulting reddening map, the grey scale reflects the variation over the field. The scale is reported at the top of the right panel.

\subsection{The spectroscopic dataset}
\label{sect_spec}

Our spectroscopic data consist of FLAMES/GIRAFFE and FLAMES/UVES data collected under the ESO program 086.D-0141 (PI: Marino). The GIRAFFE fibers were used with the HR12 setup, covering the spectral range from $\sim$5820 to $\sim$6140~\AA\ with a resolution of $\sim$18,700. The higher resolution UVES spectrograph was used in the RED 580 configuration, providing spectra centered at $\lambda$5800~\AA\ (with a spectral coverage of $\sim$2000~\AA) at a resolution of $\sim$45,000. All our target stars were observed on the same FLAMES plate for a total exposure time of 20 hours. At the wavelength of the Na D lines, the typical S/N of the fully reduced and combined GIRAFFE spectra is $\sim$100-120 and $\sim$40-80 for red and blue HB stars, respectively. UVES combined spectra have much lower S/N, ranging from $\sim$35 (red HB) to $\sim$15 (blue HB) at Na D lines. Data reduction involving bias subtraction, flat-field correction, wavelength calibration, and sky subtraction is done using the dedicated pipelines for GIRAFFE\footnote{See {\sf http://girbldrs.sourceforge.net}} and UVES (Ballester et al. 2000). 

In total we observed 108 candidate cluster members on the HB of NGC\,2808, covering the observed color range 0.05$<(B-V)<$0.90. This sample spans the HB from the reddest stars, corresponding to temperatures $\sim$ 5300-5600~K, to blue HB stars, with temperatures $\sim$9000-12000~K.

The HB of NGC\,2808 is contaminated by field stars. We determine cluster membership from radial velocities obtained using the IRAF@FXCOR task, which cross-correlates the object spectrum with a template. For the template we used a synthetic spectrum obtained through the spectral synthesis code SPECTRUM\footnote{See {\sf http://www.appstate.edu/~grayro/spectrum/spectrum.html} for more details.} (Gray \& Corbally 1994\nocite{gra94}). After applying a heliocentric correction, we find a mean radial velocity of $+$104.5~$\pm$~0.8~\kmsec\ ($\sigma$~=~8.2~\kmsec) for the whole sample. This value agrees within 3~$\sigma$ with the values in the literature (e.g.\, $+$101.6~$\pm$~0.7~\kmsec, $\sigma$~=~13.4~\kmsec, Harris 1996\nocite{har96}, 2010 edition). We then reject individual stars with values deviating by more than 3$\sigma$ from this average velocity, deeming them to be probable field stars.

Our sample of {\it bona fide} cluster stars is composed of 96 HBs: 65 distributed on the red side of the instability strip and 31 on the blue side. All the UVES targets have RVs compatible with being cluster members, providing a sample of 3 red HB and 4 blue HB of NGC\,2808 observed at higher resolution and with larger spectral coverage. Basic photometric data plus RVs for the observed stars are listed in Tab.~\ref{phot_data_tab}. Each star is flagged as red HB (RHB), blue HB (BHB), or field (F), depending on its position along the HB and its membership status. A few BHB stars have been flagged as Grundahl jump (GJ, see Grundahl et al. 1999). These GJ stars suffer from levitation of metals and sedimentation of He; consequently, their atmospheric chemical contents are not representative of the cluster abundances (see Sect.~\ref{heBHB}).

\section{Atmospheric parameters}\label{atm}

Given the large differences between red and blue HB spectra, they require different approaches to determine their atmospheric parameters. In the following Sect.~\ref{atm_bhb}, we describe the procedure for BHB stars. We discuss our analysis of RHB stars in Sect.~\ref{atm_rhb}.

\subsection{Blue Horizontal Branch}
\label{atm_bhb}

Color-temperature calibrations for BHB stars at a range of gravities are provided by the Castelli website\footnote{\sf{http://wwwuser.oat.ts.astro.it/castelli/}}. These relationships require precise photometry. The typical photometric error of our data is $\sim$0.02~mag. At 10000~K, this translates to a temperature uncertainty of $\sim$500~K. For the hottest BHB stars, with temperatures up to $\sim$12000~K, this uncertainty is as high as 1000~K. Clearly, correcting for differential reddening and zero point effects is critical for this cluster. Therefore, instead of assigning each star the temperature corresponding to its $(B-V)$ color in the color-temperature calibration, we project the target BHB stars on the ZAHB that best-fits the CMD, thus minimising the impact of the photometric uncertainties. We use the CMD in $V$ versus $(U-V)$ instead of $V$ versus $(B-V)$; with $(B-V)$ the BHB is almost vertical, and a small color error translates into a large uncertainty in atmospheric parameters.

We estimate effective temperature (\teff) and  gravity (\logg) for the BHB stars by comparing the observed CMD with theoretical models. To this aim, we considered isochrones and ZAHB loci from BaSTI\footnote{\sf{http://albione.oa-teramo.inaf.it}} (Pietrinferni et al.\ 2004, 2006). For comparison purposes we also used isochrones from PGPUC\footnote{\sf{http://www2.astro.puc.cl/pgpuc/iso.php}} database (Valcarce et al.\ 2012).

\begin{figure}
   \centering
   \includegraphics[width=8.5 cm]{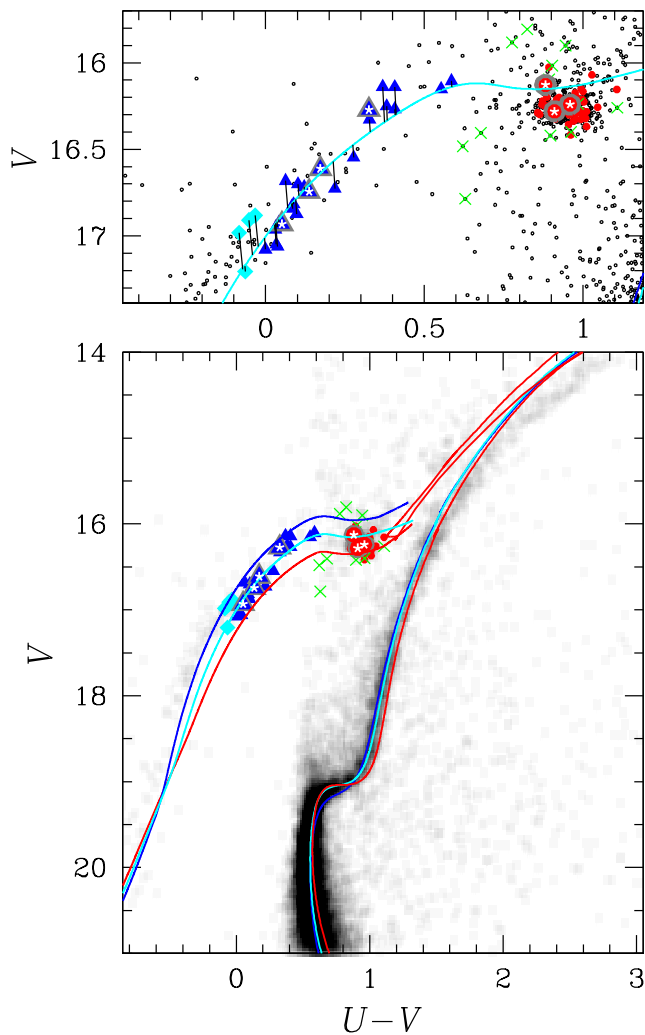}
      \caption{\textit{Lower panel:}
        $V$ versus $(U-V)$ Hess diagram for NGC\,2808.
        Spectroscopic targets are represented with different symbols:
        red circles are RHB, blue triangles represent the BHB, cyan
        diamonds Grundahl-jump stars, and green crosses probable field stars.
        The seven UVES targets are identified with grey open symbols 
        with a white star.
        The super-imposed theoretical models 
        with $Y$=0.246 (red), $Y$=0.320 (cyan), and $Y$=0.380 (blue)  
        have been extracted from the BaSTI $\alpha$-enhanced 
        isochrones with [Fe/H]=$-$1.15.
        \textit{Upper panel:} Zoom of the $V$ versus $(U-V)$ CMD in the 
        region of the HB. Black vertical lines show the projection of the BHB targets 
        on the ZAHB locus with $Y$=0.320 (see text for details).}
         \label{fig:BASTI}
   \end{figure}

In Fig.~\ref{fig:BASTI} (bottom panel) we show the $V$ versus $(U-V)$ Hess diagram for NGC\,2808 with a set of best-fitting isochrones, assuming a distance modulus $(m-M)_{V}$=15.67, reddening $E(B-V)$=0.19, [Fe/H]=$-$1.15, [$\alpha$/Fe]=$+$0.3, and age 10.0~Gyr. These values are all in agreement with estimates available in the literature (Harris\ 1996; B00; Mar{\'{\i}}n-Franch et al.\ 2009; Dotter et al.\ 2010). The red, cyan, and blue isochrones correspond to three different choices of helium, namely $Y$=0.25, $Y$=0.32, and $Y$=0.38. The positions of our spectroscopic targets on this diagram are represented with different colors and symbols, according to their position along the HB.

The region of the CMD populated by our BHB targets is best fit by assuming $Y$=0.32 (cyan model in Fig.~\ref{fig:BASTI}), and, as such, we use a ZAHB corresponding to this helium abundance to derive the atmospheric parameters for our BHB sample. We estimate effective temperatures by projecting the observed position of BHB stars in the $V$-$(U-V)$ diagram on the ZAHB locus with $Y$=0.32 (as schematically shown in Fig.~\ref{fig:BASTI}). The projection is computed as in Gallart et al. (2003, see their Sect.~4), i.e.\ by enhancing the difference in color by a factor of seven, which is determined empirically, with respect to the difference in magnitude. For a given isochrone, a star's colour is better constrained than the magnitude, as any uncertainty in distance, gravity, or reddening corresponds to a greater difference in magnitude than in colour. We obtain surface gravities from the observed $V$ magnitude and temperatures from theoretical models. We then use each star's position on the ZAHB locus to determine the corresponding stellar mass. 

An inspection of the CMD suggests that most of the BHB targets are spread around the ZAHB with $Y$=0.32. The broadening of the HB is due both to measurement uncertainties (like photometric errors, and residual differential reddening) and intrinsic spread, perhaps from star-to-star helium variations and/or the presence of stars leaving the ZAHB. Our best-fitting model used to derive the atmospheric parameters consists of only the ZAHB, while the real BHB is also populated by stars on their evolutionary tracks, with gravities lower than predicted from the ZAHB.

%
%

\begin{figure}
   \centering
   \includegraphics[width=9.4cm]{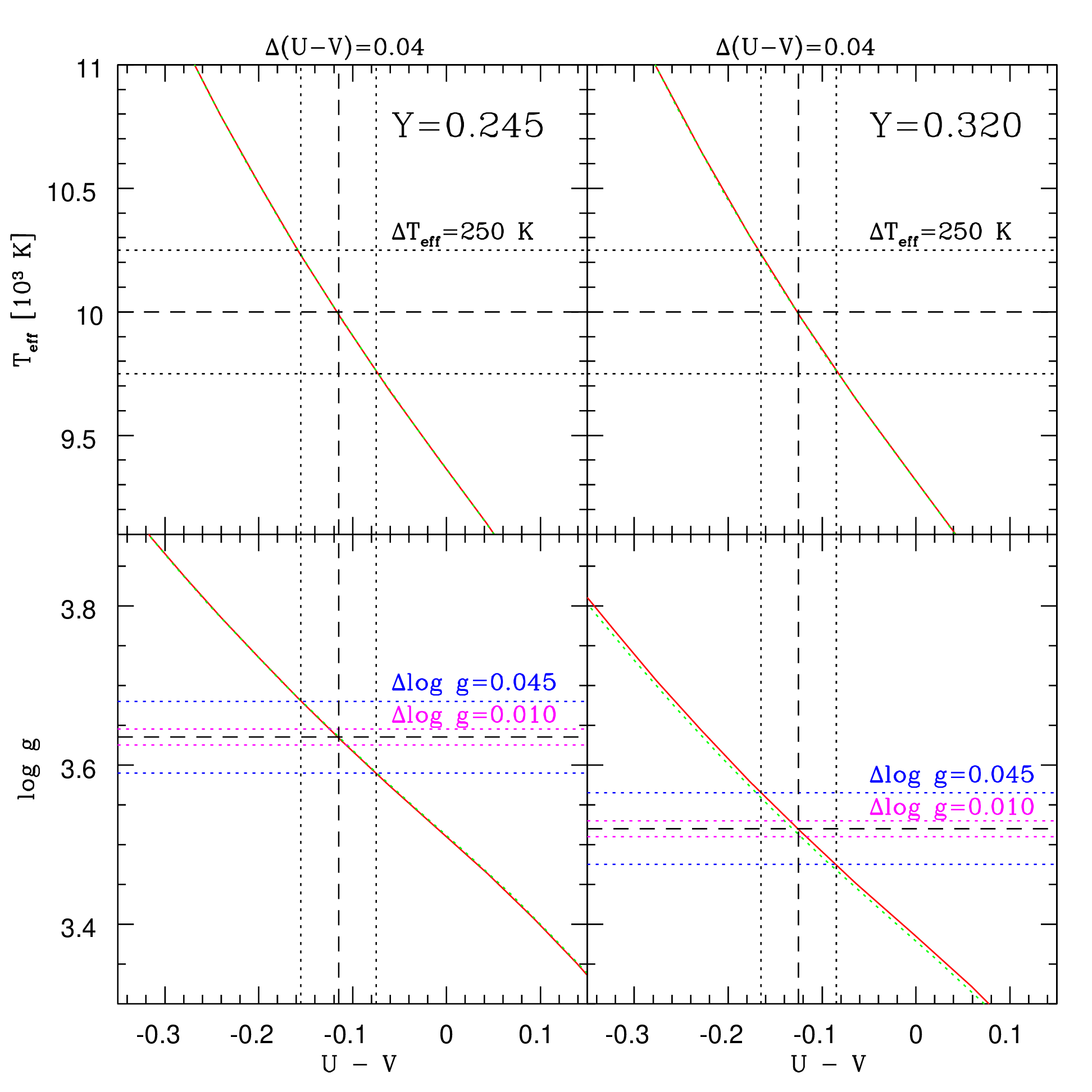}
      \caption{
ZAHB loci extracted from the PGPUC database (red lines) 
in the absolute \teff\ vs. $(U-V)$ plane
(upper panels) and in the absolute \logg\ vs. $(U-V)$ plane (bottom panels) for $Y$=0.245
(left panels) and $Y$=0.320 (right panels). 
Dashed lines correspond to 
standard BHB stars  at \teff$\sim$10000~K, and \logg=3.64~dex. 
Dotted lines correspond to an uncertainty
in $(U-V)$ of $\sim$0.04~mag.
The green dotted line is the ZAHB locus with the same properties as red lines,
except with a progenitor mass of 0.7~$M_{\odot}$.
        }
         \label{fig:PGPUC}
   \end{figure}

In Fig.~\ref{fig:PGPUC} we investigate the  impact of internal photometric errors on our temperature and gravity determinations. We have compared the ZAHB loci from the PGPUC database (red lines) in the absolute \teff-$(U-V)$ and \logg-$(U-V)$ planes for two choices of $Y$, namely $Y$=0.245 and $Y$=0.320. 
Internal photometric errors in $(U-V)$ of $\sim$0.04~mag affect the temperature determinations from isochrones by $\sim$250~K. The impact of the photometric errors on \logg\ is at most $\sim$0.05~dex.

Microturbolent velocities (\vmicro) come from the \teff-\vmicro\ empirical relation derived in Pace et al. (2006) from the least squares fit of the HB data analysed by Behr et al. (2000, 2003). For the warmer stars, which have a very stable atmosphere where helium sinks and heavy metals levitate, we adopted \vmicro=0, as suggested by Behr (2003) in NGC~288. The dispersion of \vmicro\ values around the relation found by Behr is 1~\kmsec.

From the discussion above, we consider $\pm$250~K, $\pm$0.05~dex, and $\pm$1~\kmsec\ as estimates of internal errors associated with our adopted \teff, \logg\ and \vmicro\ values for the BHB sample. To accurately determine absolute abundances for the BHB stars, we must also consider possible systematic uncertainties affecting our atmospheric parameter determinations.

Figure~\ref{fig:PGPUC} displays that a different choice of $Y$ does not affect the temperature determinations. From the theoretical point of view, models predict that HB stars near the ZAHB locus with the same $(U-V)$ colour but different helium have the same effective temperature. As such, the choice of $Y$ is not critical for the determination of \teff\ from isochrones.

In contrast, canonical models with He enhancement predict \logg\ lower than that from He-normal models. If stars on the BHB of NGC\,2808 are not He enhanced, our gravity estimates will be under-estimated by $\sim$0.15~dex. In addition, stars leaving the ZAHB track appear more luminous and have lower gravities, mimicking a ZAHB loci with higher He (of at most $Y$=0.38). When the ZAHB loci have $Y$ higher by $\sim$0.08, this changes the estimated \logg\ values by $\sim$0.12 dex (see also Moni Bidin et al.\ 2007). Age does not affect ZAHB models over the range of temperatures studied here, as suggested by the comparison between ZAHBs with a difference in age of several Gyrs (red and green dotted lines in Fig.~\ref{fig:PGPUC}).

Although our data do not allow independent measurements of atmospheric parameters to be compared with the adopted ones, we 
note the following: 
{\it (i)} on the theoretical side we do not expect significant systematics in \teff\ due to the treatment of $Y$, but we do expect systematic uncertainties in \logg\ by at most $\sim$0.15~dex; 
{\it (ii)} the best-fitted ZAHB with the BHB has a temperature of $\sim$11500~K at the so-called Grundahl jump (Grundahl et al.\ 1999), which matches the expected value for stars in this region of the HB within $\pm$500~K;
{\it (iii)} comparison of the \teff\ and \logg\ values for RHB stars from the technique described in Sect.~\ref{atm_rhb} and the methodology used for BHB stars shows agreement within 170~K in \teff\ and $\sim$0.15~dex in \logg\, with the BHB scale returning higher \teff\ and \logg\ values. Based on these arguments, we conclude that our \teff\ and \logg\ values do not seem to be affected by systematic effects larger than $\sim$200~K and $\sim$0.20~dex, respectively. The impact of these possible systematics on our results is fully discussed in Sect.~\ref{heBHB}.
 
As a final test, we verify that our choice of stellar models does not significantly change our adopted stellar parameters. We compare our adopted \teff\ and \logg\ values derived from the BaSTI database with those derived from the PGPUC databases. The two sets of atmospheric parameters show a satisfactory agreement, with mean differences of $<$\teff$_{\rm \,BaSTI}-$\teff$_{\rm \,PGPUC}>=-$23$\pm$11~K (rms=67~K) and $<$\logg$_{\rm \,BaSTI}-$\logg$_{\rm \,PGPUC}>$=+0.15$\pm$0.03~dex (rms=0.21).

\subsection{Red Horizontal Branch}
\label{atm_rhb}

The GIRAFFE HR12 setup covers a relatively small spectral range and the number of isolated Fe lines ($\sim$10 Fe\,{\sc i} and 2 Fe\,{\sc ii}) is not enough for accurate spectroscopic determination of the atmospheric parameters. To derive the atmospheric parameters of RHB targets observed with GIRAFFE, we use the Momany et al.\,(2004) photometry, corrected for differential reddening, in conjunction with $(B-V)$-\teff\ relations\footnote{We prefer not to use the 2MASS (Skrutskie et al.\ 2006) magnitudes to derive \teff\ from other colors (such as the $(V-K)$) because some of our targets are in the crowded central region of the cluster where the 2MASS magnitudes are either not available or too uncertain.} (Alonso et al. 1999, 2000). The use of these relations introduces some internal and systematic uncertainties in \teff\ due to photometric errors and reddening effects.

NGC\,2808 has an absolute reddening estimated in the literature from $E(B-V)$=0.13 (Castellani et al. 2006) to $E(B-V)$=0.23 (Piotto et al. 2002). The use of these two different absolute reddening values corresponds to a shift in the derived \teff\ of $\sim$300~K. To establish the reddening to be used in the color-\teff\ relation, and hence the \teff\ scale, we first determined atmospheric parameters for three RHB stars observed with UVES, independent of photometry. As previously mentioned, these spectra have higher resolution and a larger spectral coverage. In the typical stellar parameter space of the RHB sample, we measure $\sim$25-30 Fe\,{\sc i} and $\sim$10 Fe\,{\sc ii} lines at the S/N of our spectra, such that we can derive the atmospheric parameters from the Fe lines with sufficient confidence. We determine \teff\ by imposing the excitation potential equilibrium of the Fe\,{\sc i} lines and gravity with the ionisation equilibrium between Fe\,{\sc i} and Fe\,{\sc ii} lines. Note that for \logg\, we impose Fe\,{\sc ii} abundances that are slightly higher (by 0.07-0.08 dex) than the Fe\,{\sc i} ones to adjust for NLTE effects (Lind et al. 2012; Bergemann et al. 2012). For this analysis, microturbolent velocities, \vmicro\, were set to minimize any dependence on Fe\,{\sc i} abundances as a function of EWs.

In Tab.~\ref{uves_atm} we compare effective temperatures for UVES RHB stars derived from spectroscopy and those obtained from the Alonso $(B-V)$-\teff\ calibrations, assuming a reddening of $E(B-V)$=0.19. We find that this reddening value best reproduces the spectroscopic \teff\ measurements listed in Tab.~\ref{uves_atm}, with two stars having the same \teff\ within $\sim$30~K, and one within $\sim$100~K. This value of $E(B-V)$ corresponds with that provided by B00 and agrees within errors with the value determined from RR-Lyrae by Kunder et al. (2013, $E(B-V)$=0.17$\pm$0.02). A reddening of $E(B-V)$=0.22 (as listed in the Harris catalog) gives too high a \teff\ with respect to the spectroscopic values. 

The approach adopted for UVES spectra, independent of photometry, allows us to evaluate systematic differences with the atmospheric parameters determined for the GIRAFFE targets via the color-\teff\ calibrations, and adopt the same \teff\ scale for UVES and GIRAFFE targets. Of course, un-accounted sources of systematic errors, such as the LTE approximation, could exist in the spectroscopic determination of \teff. However, as our \teff\ scale agrees with recent reddening estimates for the cluster, we are confident in our adopted temperature scale. 

In our study of RHB stars we are mainly interested in the star-to-star internal abundance variations. We are thus sensitive to the internal errors associated with the atmospheric parameters. Internal uncertainties on the \teff\ derived from the Alonso et al. (1999) calibrations are mainly due to photometric errors and differential reddening, which is expected to be a significant effect for a cluster with reddening comparable to NGC\,2808. Indeed, the map of Fig.~\ref{fig:comp} shows that star-to-star color variations in the 11.25\arcmin$\times$11.25\arcmin\ field around the cluster center can be as large as $\sim$0.06 mag, which corresponds to variations in \teff\ of $\sim$200~K. To increase the precision on the temperature estimate, we correct the photometry for these effects. After this correction, for bright HB stars the internal errors associated with our $(B-V)$ colors are typically around 0.02~mag, corresponding to errors in \teff\ of the order of $\sim$50-60~K. 

As a check we re-determine temperatures using the B00 photometry. The mean difference between the \teff\ derived from the two different photometries is 75$\pm$5~K (rms=37~K, with the B00 \teff\ lower). Comparison of the two different photometric catalogs suggests that, after the absolute reddening value for the cluster has been fixed, we expect an accuracy for the RHB \teff\ of $\sim$100~K. Uncertainty due to internal photometric errors, as well as the rms associated with the mean difference between the \teff\ derived from the two photometric data sets, is thus less than this value. This suggests that internal errors associated with our photometric \teff\ are no larger than $\sim$50~K.

Surface gravities for the GIRAFFE RHB stars are obtained from the apparent $V$ magnitudes from Momany et al.\ (2004), corrected for differential reddening, the \teff,\ bolometric corrections from Alonso \etal\ (1999)\nocite{al99}, and an apparent distance modulus of $(m-M)_{V}$~=15.67, which is obtained from isochrone fitting and consistent with the value obtained in B00. We assume masses taken from isochrones, which range over an interval of $\sim$0.1~$M_{\odot}$ (0.61$\lesssim$$M$$\lesssim$0.75 $M_{\odot}$). Our \logg\ determinations for stars observed with GIRAFFE are affected by intrinsic uncertainty in mass owing to stochastic mass loss in the RGB phase. Internal errors in \teff\ values of $\sim \pm$50~K and of $\sim \pm$0.1 in mass, affect the \logg\ values by $\sim \pm$0.02 and $\pm$0.06 dex, respectively. The internal photometric uncertainty associated with our $V$ mag modifies our \logg\ by $\sim$0.01 dex. All these effects, added in quadrature, contribute to an internal error in \logg\ $\lesssim$0.10~dex.

Microturbulent velocities for GIRAFFE data cannot be independently determined from the spectral lines or photometry. We utilise a \vmicro-\teff-\logg-metallicity relation developed for the Gaia-ESO survey (GES, Gilmore et al. 2012) which was obtained from different literature sources\footnote{{\sf http://great.ast.cam.ac.uk/GESwiki/GesWg/GesWg11/Microturbulence}}. For this relation, we adopt a mean [Fe/H]=$-$1.14 (Harris catalog), and our estimates of \teff\ and \logg\ explained above. The range in our estimated \vmicro\ is quite small, from 1.50 to 1.56~\kmsec, with a mean value of 1.54~\kmsec\ ($\sigma$=0.01). This range is much smaller than that observed for the UVES RHB stars (see Tab.~\ref{uves_atm}). We note that this relation for \vmicro\ has not been calibrated for HB stars and is formally applicable only to giants and dwarfs. However, comparison with \vmicro\ values derived from UVES spectra using Fe\,{\sc i} spectral lines, suggests our values are accurate to $\sim$0.2~\kmsec.

As a further test, we derive \vmicro\ values from Fe\,{\sc i} spectral lines from GIRAFFE data, despite the small number of features available. We obtain a median value of \vmicro=1.64$\pm$0.13~\kmsec. We cannot establish if our sample of stars have an intrinsic dispersion in \vmicro, or if this higher dispersion obtained from spectroscopy is due to observational uncertainties. Given the limited number of spectral lines in the GIRAFFE data, we use the \vmicro\ values from the GES empirical relations for our analysis. We also assume an internal error of 0.15~\kmsec, similar to the dispersion associated with spectroscopic \vmicro.

\subsection{Tests on the atmospheric parameters}

We test our adopted atmospheric parameters for RHB stars by comparing them with those available in the literature, and those obtained with different techniques.

Gratton et al.\ (2011; hereafter G11) analysed GIRAFFE spectra of red and blue HB stars in NGC\,2808. These authors estimated effective temperatures of RHB stars from their $(B-V)$ and $(V-K)$ colors (with higher weight to $(B-V)$), using the calibration of Alonso et al.\ (1999). For all of the stars, they assumed the average reddening of NGC\,2808 given by Harris (1996) and the  $E(B-V)/E(V-K)$ value from Cardelli et al.\ (1989). Gravities were estimated from masses, luminosities, and effective temperature by assuming a constant mass of 0.7$M_{\rm \odot}$ for RHB stars. Metallicity values are obtained from the analysis of Fe\,{\sc i} lines. For further details, we refer the reader to Sect.~3 of G11.

\begin{figure}
   \centering
   \includegraphics[width=7. cm]{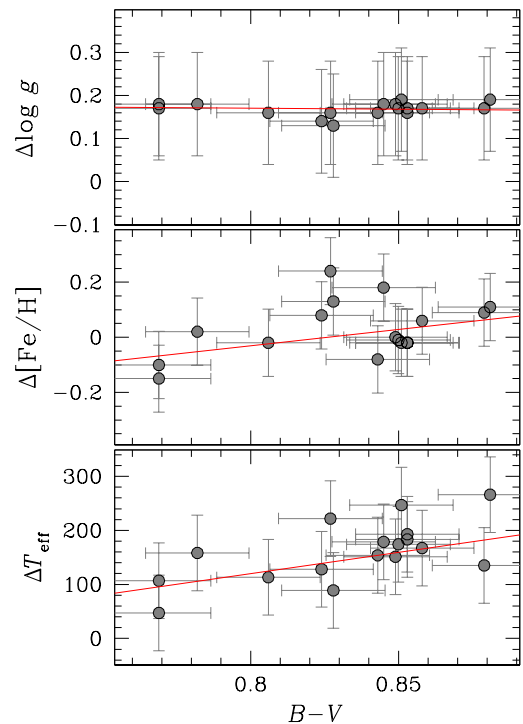}
      \caption{ 
Upper, middle, and lower panels show the difference between gravity
($\Delta$\logg), iron ($\Delta$[Fe/H]), and effective temperature
($\Delta$\teff) measurements derived by 
G11 and those of this paper as a function of the $(B-V)$ color. 
Red lines are the best-fit straight lines.}
   \label{compTG}
   \end{figure}

Seventeen RHB stars analysed by G11 are included in our work. We compare the atmospheric parameters obtained in these two works in Fig.~\ref{compTG}, where we show $\Delta$\teff=\teff$_{\rm G11}-$\teff$_{\rm this\  paper}$ (lower panel), $\Delta \rm [Fe/H]=[Fe/H]_{\rm  G11}-[Fe/H]_{\rm this\ paper}$ (middle panel), and $\Delta\logg=\logg_{\rm G11}-\logg_{\rm this\ paper}$ (upper panel) as a function of the $(B-V)$ color corrected for differential reddening.
 
On average, \teff\ estimates by G11 are systematically higher than those measured in our paper by 160~K with a scatter of $\sim$55~K. This offset is in part due to the different $E(B-V)$ we use with the Alonso et al. (1999) calibration, as our value provides \teff\ that are lower by $\sim$100~K. The temperature difference is mildly correlated with the color, with a Spearman's coefficient equal to 0.45. This may be due to our corrections in $(B-V)$ for differential reddening, zero point effects, and/or the larger photometric uncertainties and reddening variations in $(V-K)$.

Iron abundances from G11 and from this paper differ by 0.03~dex with a scatter of 0.1~dex. There is a mild trend between $\Delta$[Fe/H] and the color, which is a consequence of the \teff-$(B-V)$ trend. The difference in [Fe/H] ranges from $\sim -$0.1~dex for stars with $(B-V) \sim$0.77 to $\sim +$0.1~dex for stars with $(B-V) \sim$0.88.    

Gravities from G11 are systematically higher by $\sim$0.17~dex. This difference appears to be constant over the analysed color interval, and could be due, in part, to the assumed distance modulus, the \teff\ scale, and the different choice for the stellar masses. Indeed, the distance modulus, used in G11, is lower ($(m-M)_{V}$~=~15.59) than the value adopted here and, as such, gives higher \logg\ values by 0.08~dex, on average. A systematic error in temperature of 160~K will change gravities by another 0.06~dex. Finally, rather than a fixed mass of 0.7~$M_{\rm \odot}$ as in G11, we use masses from isochrones that range towards lower stellar mass values.

\begin{figure}
   \centering
   \includegraphics[width=10cm]{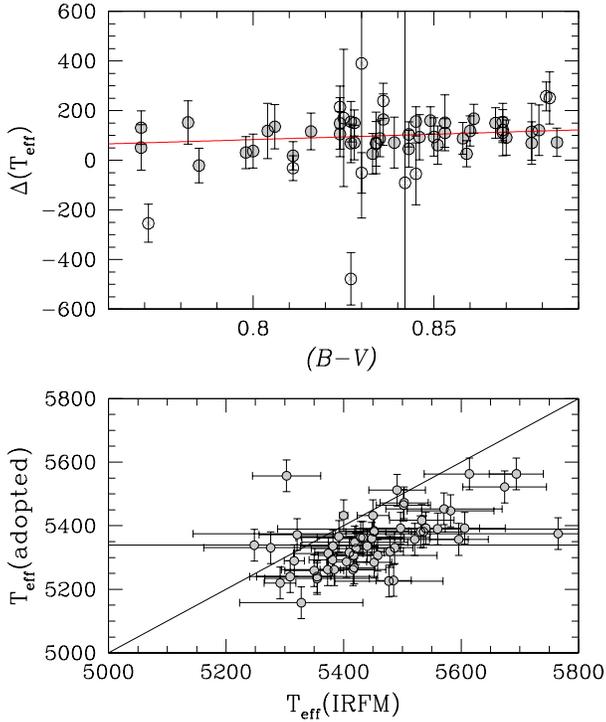}
      \caption{{\it Lower panel}: adopted \teff\ as a function of \teff\
   derived from the IRFM. 
{\it Upper panel}: difference between \teff\ estimates from the IRFM 
and the adopted values as a function of the $(B-V)$ color. The empty circles are 2~$\sigma$ outliers and have been rejected from the least squares fit. }
   \label{fig:irfm}
   \end{figure}

As a further test of our effective temperatures, we compute \teff\ values from the infrared flux method (IRFM, see Casagrande et al.\ 2010), using our $B$, $V$, $JHK$ from 2MASS, and our adopted \logg. Figure~\ref{fig:irfm} shows the comparison between the \teff\ values adopted here with those derived from the IRFM (\teff(IRFM)). The error bars associated with the \teff(IRFM) are set equal to the scatter in temperature derived from $J$,$H$, and $K$. Additional uncertainties in these temperatures come from the errors associated with the adopted \logg, metallicities, zero-points in the \teff\ scale, and reddening effects. The \teff\ from the IRFM (listed in the last column of Tab.~\ref{gir_par_tab}) are systematically higher by 87$\pm$16~K, with a rms of 121~K. The difference between the two \teff\ scales may be caused by an offset in reddening. The contribution to the \teff(IRFM) error introduced by the adopted \logg\ is negligible, as a relatively large error in \logg\ of $\sim$0.5~dex, corresponds to an error in \teff\ of $\lesssim$50~K. As shown in the upper panel of Fig.~\ref{fig:irfm}, there is no significant trend between the difference $\Delta$\teff=\teff(IRFM)-\teff$_{\rm this\  paper}$ with the $(B-V)$ color.
 
%
%

%
%

\section{Chemical abundances}
\label{abunds}

Chemical abundances for all elements, with the exception of helium, were obtained from the equivalent widths (EWs) of isolated spectral lines, whose profiles have been fit with a Gaussian. The spectra of RHB stars are very different from those of BHB stars. 
For RHB stars, in the GIRAFFE HR12 spectral range, we measure Fe\,{\sc i} and Fe\,{\sc ii} from 6-8 and 1-2 spectral lines, respectively. For our measurements of Na, Si, Ca, Ti, Mn, and Ba, there are one to two transitions for each element. The UVES spectra have larger wavelength coverage and higher resolution; thus, we can determine chemical abundances in the RHB stars for additional elements, i.e.  Fe, $p$-capture elements Na, Mg, $\alpha$-elements Si, Ca, Ti, iron-peak elements Sc Cr, Ni, Zn, and $n$-capture elements Y, Ba and Nd. 
In HR12 observations of cooler BHB stars, we are able to measure only the Na resonance doublet. Thus, for the BHB stars with \teff$\gtrsim$9000~K, we cannot derive reliable Na abundances because of the very weak Na lines. For hotter BHB stars, the He feature at $\sim$5875.6~\AA\ is detectable in the GIRAFFE spectra. For the four BHB observed with UVES, we determine iron abundances from 6-7 Fe\,{\sc ii} lines, and magnesium from the lines $\lambda$5173~\AA\ and $\lambda$5184~\AA. Given the low S/N of the UVES spectra, the Na D lines are not well detected, and we are able to estimate only upper limits to the Na content.

Our atomic data is from the NIST{\footnote{\sf http://physics.nist.gov/PhysRefData/ASD/lines\_form.html}} (National Institute of Standards and Technology) compilation. For the few lines for which NIST does not provide the oscillator strengths (log~$gf$), we use  log~$gf$ from VALD (Kupka 2000) and Mel{\'e}ndez \& Barbuy (2009) for Fe\,{\sc ii} lines.

We derive chemical abundances from a local thermodynamic equilibrium (LTE) analysis using the latest version of the spectral analysis code MOOG (Sneden 1973). For the helium line analysis we employ a hybrid NLTE approach as described by Przybilla et al. (2006, see Sect.~\ref{bhbpar} for further details). For both RHB and BHB stars we apply NLTE corrections to Na abundances, as prescribed by Lind et al. (2011).

\subsection{Red Horizontal Branch stars}
\label{rhb}

\subsubsection{UVES sample}
\label{uves_rhb}

The derived chemical abundances for the three RHB stars with UVES spectra are listed in Tab.~\ref{uves_abb}. For each element, we report the abundance relative to Fe\,{\sc i} or Fe\,{\sc ii} according to their ionisation state; this minimises the dependence of the derived abundance on the model atmosphere. 

To determine the influence of atmospheric parameter uncertainties on our chemical contents, we re-derive abundances changing \teff/\logg/\vmicro/[A/H] by $\pm$100~K/$\pm$0.15/$\pm$0.20~\kmsec/0.10~dex, respectively. Neutral iron is primarily sensitive to \teff\ and \vmicro; it increases by 0.09~dex with a larger temperature and decreases by 0.06~dex with a larger microturbulence. As expected, \logg\ has the largest effect on singly ionised iron; a larger gravity value increases the measured abundances by 0.06~dex. In most cases, the corresponding variations in the abundances relative to Fe are small (a few hundredths of a dex).

Changes in \teff\ by $\pm$100~K results in the largest variations in [Si, Ba, Nd/Fe]; these abundances change by  $\mp$0.06, $\pm$0.07, and $\pm$0.05 dex, respectively. Barium abundances are also sensitive to \vmicro. Variation in this parameter by $\pm$0.20~\kmsec\ corresponds to a change in [Ba/Fe] by $\mp$0.12~dex. The other parameters have small impact on the derived abundances for these elements over Fe.

The analysed elements show one of the typical chemical patterns of a GC, e.g.\ $\alpha$ enhancement indicated by the abundances of Si, Ca, Ti. Mg is not considered to be a pure $\alpha$ element, as it is affected by the $p$-capture reactions that convert Mg to Al. Unfortunately, the quality of our spectra is not good enough to infer Al abundances, however the Mg content appears to be un-depleted in all three stars.

We determine the sodium abundance from both the resonance doublet at $\sim$5890~\AA\, and the feature at $\sim$5688~\AA. The resonance doublet is heavily affected by deviations from LTE. We apply the NLTE corrections from Lind et al. (2011) to all our Na spectral lines, significantly decreasing the abundance from the Na resonance doublet by $\sim$0.4-0.5~dex. The correction for the other Na feature is smaller, $\sim$0.10~dex. After the NLTE correction, the Na abundances from all three lines are consistent within errors, and the standard deviation associated with the mean Na value of each star decreases, except for star \#21854, for which the standard deviation slightly increases. The agreement between the NLTE-corrected abundances derived from the Na resonance doublet and the 5688~\AA\, line suggests that, with the proper NLTE corrections, abundances derived from the very strong Na resonance lines are reliable.

\subsubsection{GIRAFFE sample}
\label{giraffe_rhb}

In Tab.~\ref{gir_par_tab} we list the adopted atmospheric parameters and corresponding abundances determined from the GIRAFFE spectra of the RHB stars. The only feature available to determine Na abundances in this data is the resonance doublet. Results obtained on UVES data suggest that we can confidently use these strong lines for abundance determinations, once NLTE corrections have been applied. As done for the UVES data, we apply NLTE corrections from Lind et al.\ (2011), decreasing the mean [Na/Fe] content for our RHB sample from [Na/Fe]$_{\rm LTE}$=0.48$\pm$0.02 ($\sigma$=0.15) to [Na/Fe]$_{\rm NLTE}$=0.09$\pm$0.02 ($\sigma$=0.15). This correction decreases the mean difference between the abundances given by the two single lines of the doublet from [Na/Fe]$_{\rm 5889-5895}$=$-$0.09 to [Na/Fe]$_{\rm 5889-5895}$=0.01~dex, indicating that accounting for NLTE effects improves our Na values.

The mean abundance ratios relative to Fe for the other elements measured from GIRAFFE spectra of RHB stars are:
 [Si/Fe]{\sc i}=0.44$\pm$0.01 ($\sigma$=0.09); 
 [Ca/Fe]{\sc i}=0.52$\pm$0.01 ($\sigma$=0.09);  
 [Ti/Fe]{\sc i}=0.21$\pm$0.02 ($\sigma$=0.12); 
 [Mn/Fe]{\sc i}=$-$0.42$\pm$0.01 ($\sigma$=0.10); 
 [Fe/H]{\sc i}=$-$1.22$\pm$0.01 ($\sigma$=0.07); 
 [Fe/H]{\sc ii}=$-$1.10$\pm$0.01 ($\sigma$=0.08); 
 [Ba/Fe]{\sc ii}=0.14$\pm$0.02 ($\sigma$=0.19).
We notice that these mean chemical abundances are similar to those observed in other clusters. Silicon and titanium are enhanced, reflecting the $\alpha$-enhancement observed in field and GCs stars at similar metallicities, and manganese is sub-solar, with a mean value consistent with that observed in other clusters (Sobeck et al. 2006). The quite high average value for [Ca/Fe] is due to the limited spectral features available; we have only the transition $\lambda$6122\AA\, which is on the flat part of the curve of growth. As a consequence, it returns abundances systematically higher by $\sim$0.20~dex than other transitions, as verified on UVES spectra. 

To derive estimates of the internal uncertainties associated with our determinations, we calculate the sensitivity of the abundances to various sources of internal errors, e.g.\ uncertainties in the model atmosphere and in the EWs. The stellar parameter internal uncertainties are (\teff/\logg/[Fe/H]/\vmicro): $\pm$50~K/0.1~dex/0.07~dex/0.15~\kmsec\ (see Sect.~\ref{atm_rhb}). We set the uncertainty associated with the [Fe/H] values equal to the dispersion of the obtained Fe\,{\sc i} abundances, assuming that the cluster is mono-metallic. To quantify the effect on the abundances, we run a series of models in which each parameter is varied by its corresponding uncertainty one at a time.  

The contribution to the errors given by EW uncertainties has been calculated by varying the EWs of spectral lines by $\pm$4 m\AA. This is the typical error associated with our EW measurements, determined by comparing EWs for stars with similar atmospheric parameters and abundances. The variations in the Fe\,{\sc i} abundances from EW uncertainty are then divided by the square root of the number of available spectral lines. Hence, since the EW measurement errors are random, the uncertainty is lower for those elements with a large number of lines. For the other elements we have only one or two lines, and the error contribution by EW uncertainties is higher. Given the importance of the Na abundances for the discussion (see Sect.~\ref{results}), we perform an additional test to estimate the error introduced by uncertainties in EW, measuring the Na EWs from each single exposure. The associated rms/$\sqrt{N-1}$ (N=number of exposures) is $\sim$4~m\AA, agreeing well with the EWs error estimate used to determine the abundance uncertainties earlier. 

The variations in abundances obtained by varying the atmospheric parameters and EWs are listed in Tab.~\ref{gir_errors}. In this table we also list the squared sum of these different contributions ($\sigma_{\rm total}$), and the observed dispersion ($\sigma_{\rm obs}$) for each element. In some cases, the dispersions are large, as we expect from an analysis limited to only one or two spectral lines. The expected error values $\sigma_{\rm total}$ are only rough estimates of the internal uncertainties we expect for our abundances, but they suggest that, in our analysis, the main contributors to the errors are EW uncertainties. Overall, the observed dispersions agree well with the expected values, apart for Fe\,{\sc ii}, where the expected errors are slightly over-estimated, and Na, where the observed dispersion is more than twice as large as expected. We discuss the Na spread further in Sect.~\ref{results}.

\subsection{Blue Horizontal Branch  stars}
\label{bhbpar}

\subsubsection{UVES sample}
\label{uves_bhb}

The UVES sample of BHB stars consists of four stars, with 9200$<$\teff$<$11000~K. Despite the larger available spectral range, the only measurable iron lines are 6-7 singly ionised lines. Hence, as with the GIRAFFE data, we use \teff\ and \logg\ values from isochrones, as explained in Sect.~\ref{atm_bhb}. This assures uniformity in the \teff\ and \logg\ scales used for UVES and GIRAFFE BHB stars.

Atmospheric parameters and chemical abundances obtained for these stars are listed in Tab.~\ref{BHB_TAB-UVES}. Given that our information on the metallicity of these stars comes only from singly ionised Fe lines, we choose [Fe/H]{\sc ii}) as our metallicity estimate. The NLTE effects for these singly ionised lines are smaller (Lind et al. 2012; Bergemann et al. 2012). We are able to infer the Na content for only the coldest star. For the other three, we provide an upper limit to the Na abundances. 

We estimate the internal abundance errors for these stars by altering the atmospheric parameters, changing \teff\ by 250~K and \vmicro\ by 1~\kmsec, and redetermining the abundances. Iron content derived from singly ionised lines increases by $\sim$0.06~dex with increasing \teff\ and by $\sim$0.12~dex with decreasing \vmicro. A variation in \teff\ by $\pm$250~K changes magnesium and sodium abundances relative to Fe\,{\sc ii} by $\pm$0.12~dex, and $\pm$0.10~dex, respectively, while a $\pm$1~\kmsec\ change in \vmicro\ varies both these abundances by roughly $\pm$0.10~dex. Formally, \logg\ is not affected by high internal errors (Sect.~\ref{atm_bhb}), and should not significantly contribute to the internal abundance uncertainties. 
Systematics in \logg\ of $\pm$0.20~dex will change Fe\,{\sc ii} by $\sim \pm$0.05~dex, and Na and Mg relative to Fe\,{\sc ii} by $\mp$0.10~dex. 
In addition to uncertainties from the model atmosphere, the relatively low S/N of these spectra translates into a non-negligible error contribution ($\sim$0.10~dex for all the species) from EW measurement errors to the {\it real} abundance uncertainties.

From the analysis of 12 RGB stars in NGC\,2808, Carretta et al. (2009) found three of them with under-solar [Mg/Fe] abundance. The low Mg abundances in these three stars is accompanied by a higher Na abundance, typical of second generation(s) stars. None of our 4 BHB stars analysed with UVES shows a similar low Mg abundance, but the [Na/Fe] abundance, measured for just one of the UVES sample, resembles the values of the Na-rich stars observed on the RGB. Given the large internal uncertainties associated with our [Mg/Fe] measurements, that are $\sim$0.20~dex, it is very difficult to draw firm conclusions here on the basis of our results on a sample of four stars. We can just not exclude that these stars, as suggested by theoretical studies of the HB (D'Antona et al. 2005), may not be the counterpart of the extremely Na-rich (O and Mg poor) stars observed on the RGB, but an intermediate population with mild Mg-depletion. Three out four BHB stars in our sample show in fact slightly lower Mg abundances than the RHB stars.

\subsubsection{GIRAFFE sample}
\label{giraffe_bhb}

The chemical abundances for Na and He obtained for the BHB stars are listed in Tab.~\ref{gir_par_tab} and Tab.~\ref{tab:bhb_he}, respectively. For the hotter stars in our sample (21 stars in total with \teff$>$9000~K) we are able to detect the He\,{\sc i} $\lambda$5875.6\,{\AA} line, which has been used in previous studies (e.g.\ Behr\ 2003; Villanova et al.\ 2009, 2011).

\begin{figure}
   \centering
   \includegraphics[width=8.5 cm]{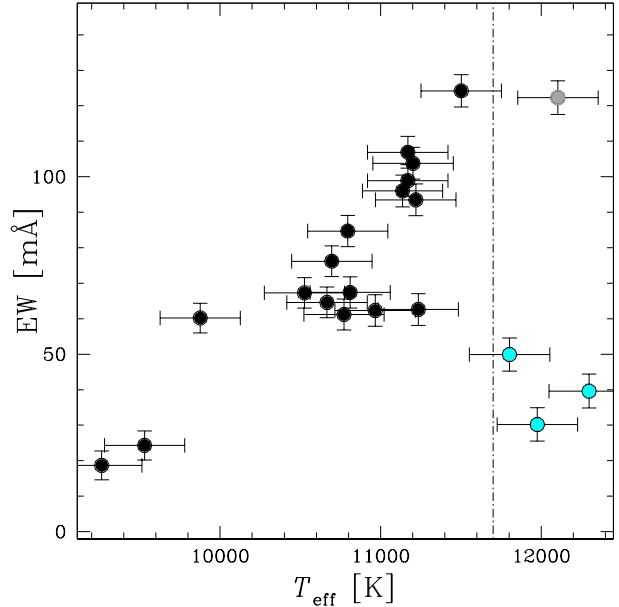}
      \caption{
        Equivalent widths of the He line as a function of \teff.
        Stars with \teff$\gtrsim$11500~K, likely affected by helium 
        sedimentation, are represented by cyan circles. One of these 
        stars, shown in grey, does not exhibit
        the extreme drop in the EW of the He line shown by 
        the other three stars at similar temperature.
        }
        \label{he}
   \end{figure}

In Fig.~\ref{he} we illustrate the behaviour of the EWs of the He spectral line with effective temperature. The EW steeply increases from $\sim$10 m\AA\, up to $\sim$130 m\AA\, over the \teff\ interval $\sim$9000-11500 K, and then falls down to $\sim$30-50 m\AA\ at higher temperature. This drop is likely a consequence of helium settling and metal levitation, which results in a depletion of the surface abundance of He and an overabundance of some heavy elements (Behr 2003; Fabbian et al.\ 2005). Since these phenomena cause the surface abundances to dramatically deviate from those of the star as a whole, we exclude from the following discussion the three stars with \teff$>$11500 K and anomalous values of the helium-line EW (stars represented in cyan in Fig.~\ref{he}). We also exclude one star (\#4129, shown in grey in Fig.~\ref{he}) with a temperature hotter than the GJ that does not show an extreme drop in the EW. On the basis of its iron abundance (Pace et al.\ 2006), this object appears to be suffering from mild metal levitation, as its metallicity is higher than the cluster mean metallicity, but sensibly lower than that of hotter GJ stars (see Tab.~3 in Pace et al.\ 2006).

\begin{figure*}
\centering
\includegraphics[width=15cm]{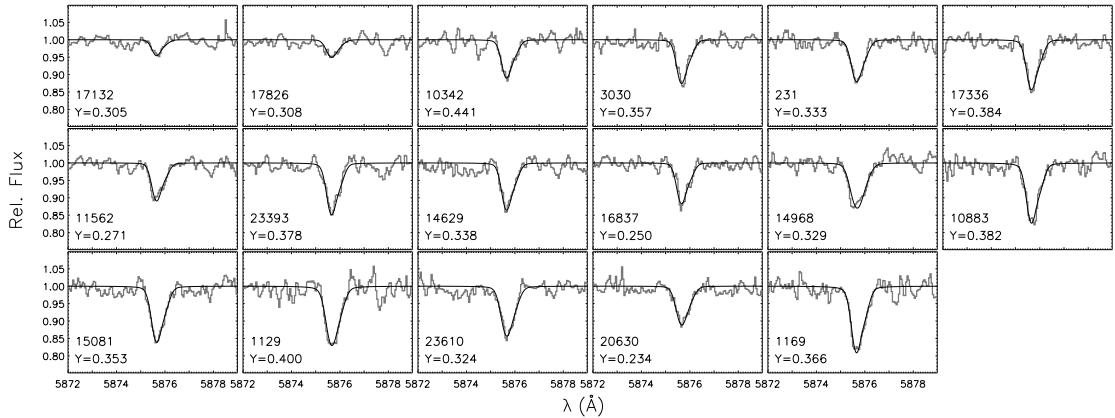}
\caption{Comparison of synthetic NLTE spectra (black lines)
with observed spectra (grey lines) for 17 BHB stars cooler than the
Grundahl jump. Stellar identifications and the best-fit helium mass
fractions $Y$ are indicated. The individual panels are ordered by
increasing effective temperature of the stars.}
\label{He-fits}
\end{figure*}

\begin{figure*}
\includegraphics[width=15cm]{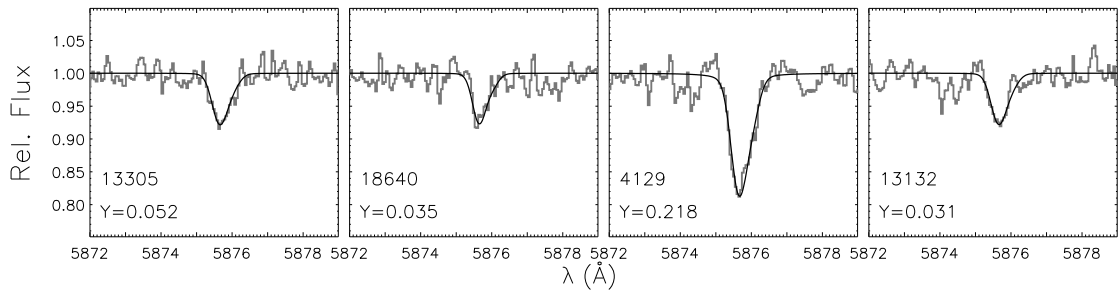}
\caption{As Fig.~\ref{He-fits}, but for the four sample stars beyond
the Grundahl jump.}
\label{HeGJ-fits}
\end{figure*}

To estimate the abundance of He, we perform a spectral synthesis of the observed line $\lambda$5875\AA. The best-fit to observations are shown in Fig.~\ref{He-fits} and Fig.~\ref{HeGJ-fits} for BHB stars with temperatures lower and higher than \teff=11500~K, respectively. For ten stars in the sample, we must account for a non-zero projected rotational velocity (\vsini) to match the observed He line profile (see Tab.~\ref{tab:bhb_he}). The adopted values agree with the \vsini\ range derived for BHB stars in NGC\,2808 by Recio Blanco et al. (2004, 5$\lesssim$\vsini$\lesssim$15~\kmsec). The synthetic spectra, for the $T_\mathrm{eff}$/$\log g$/$\xi_\mathrm{t}$ values given by the isochrones and metallicity equal to $-$1.15 (see Sect.~\ref{atm_bhb}), were calculated employing NLTE line-formation computations with DETAIL/SURFACE (Giddings 1981; Butler \& Giddings 1985) on ATLAS12 LTE model atmospheres, using the He\,{\sc i} model atom of Przybilla (2005). The most relevant extension here is the implementation of opacity sampling (instead of opacity distribution functions) to account for the line blocking, adopting the technique and data discussed by Kurucz (1996). The approach has been successfully applied to spectrum synthesis analyses of Population\,II BHB and blue straggler stars before (Przybilla et al. 2010; Tillich et al. 2010).

%
%

%
%

%
%

Uncertainties associated with the He abundances were determined for seven BHB stars (\#17132, \#10342, \#3030, \#14968, \#1169, and GJ stars \#13305 and \#4129), at characteristic points of the parameter range. Considering typical uncertainties in \teff/\logg/\vmicro/[$A/H$] of 250\,K/0.05\,dex/1.00\,\kmsec/0.07\,dex, and those due to continuum placement, we find a {\em typical} total uncertainty (sum over the squared contributions) of 0.05-0.06 and 0.02-0.03 in mass fraction for stars below and above the Grundahl jump. The main source of uncertainty stems from inaccuracies in the \teff-estimation; all other factors are relatively minor.

\begin{figure}
\includegraphics[width=8.5cm]{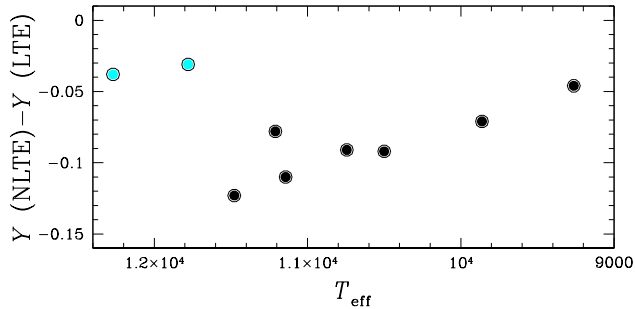}
\caption{Difference between NLTE and LTE helium mass fractions as a function of the effective temperature. Colors are the same as in Figure\,\ref{he}.}
\label{DiffNLTE}
\end{figure}

We find that NLTE effects on the helium abundance estimates are substantial. The difference between LTE and NLTE helium abundances depends on both the atmospheric parameters and helium content of the star. This difference is plotted in Fig.~\ref{DiffNLTE} as a function of \teff\ for a representative sample of stars. It varies between between 0.03 to 0.12 in mass fraction from star to star, with higher values derived for LTE. The NLTE effects on He\,{\sc i} $\lambda$5875.6\,{\AA} tend to increase when the line gets stronger, as the feature is not saturated in our hotter BHB stars. This results from a slight NLTE overpopulation of the lower level $2p$\,$^3$P$^o$ and a slight NLTE underpopulation of the upper level $3d$\,$^3$D, both by less than 5\% relative to detailed equilibrium values throughout the line-formation region. Both levels play an important role in the recombination cascade, which leads to the NLTE overpopulation of the metastable lowest state in the helium triplet spin system. Neglect of NLTE effects systematically biases the He abundances, exceeding the combined effect of all other sources of uncertainty (internal and external).

We do not expect any significant uncertainty due to inaccuracies in the atomic data employed for the model atom construction of He\,{\sc i}. The atomic data for He\,{\sc i} measured in experiments and provided by ab-initio computations has the highest accuracy and precision next to that of hydrogen. Synthetic spectra obtained with the He\,{\sc i} model atom adopted here compare well to those obtained with other NLTE codes (see e.g. Nieva \& Przybilla 2007; Przybilla et al. 2011). Moreover, and even more importantly, this model atom can reproduce the {\em entire} observed visual/near-IR He\,{\sc i} line spectra in a large variety of star types: in supergiants with similar $T_\mathrm{eff}$ than the BHB stars investigated here (Przybilla et al. 2006a), in B-type main sequence stars (Nieva \& Przybilla 2012), in extreme horizontal branch stars (Przybilla et al. 2006b) and in extreme helium stars (Przybilla et al. 2005).

%
%

\section{The abundance distribution of stars along the horizontal branch}
\label{results}

\begin{figure}
   \centering
   \includegraphics[width=7.8cm]{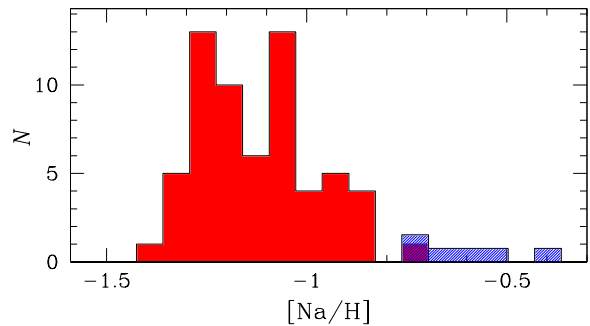}
      \caption{Histogram distribution of [Na/H] for the HB stars in NGC\,2808. Red and blue HB stars have been indicated with red and blue colours, 
   respectively.}
        \label{fig:NaDistr}
   \end{figure}

The distribution of [Na/H] for HB stars observed with GIRAFFE is shown in Fig~\ref{fig:NaDistr}. There is a large star-to-star variation in sodium abundance, with [Na/H] ranging from $\sim -$1.4 up to $\sim -$0.5~dex. In this section, we investigate the relation between the sodium abundance of a star and its position along the HB. To this aim, we consider two groups of blue and red HB stars, represented with different colors in Fig~\ref{fig:NaDistr}.

Following B00, the BHB is formed by three segments separated by different gaps: extended blue tail 1, 2 and 3 (EBT1, EBT2 and EBT3) in order of increasing temperature. Specifically, the EBT1 includes hot HB stars located between the blue edge of the RR-Lyrae instability strip and the first gap; the EBT2 includes extreme HB stars located between the first and the second gap; and stars in the EBT3 are extreme HB stars hotter than the second gap.

All our analysed BHB stars (with the exception of Grundahl jump stars) belong to the EBT1. The Na richest stars (with [Na/H]$\gtrsim -$0.8) are EBT1 stars, while stars on the red side of the instability gap show lower sodium values. This distinction suggests that the HB morphology in GCs is closely connected with the present stellar populations and demonstrates that the EBT1 of NGC\,2808 is made of second-generation stars. In the following subsection, we discuss the Na abundance distribution for RHB stars.

%
%

\subsection{The sodium content of red horizontal branch stars\label{narhb}}

\begin{figure}
   \centering
   \includegraphics[width=8.55cm]{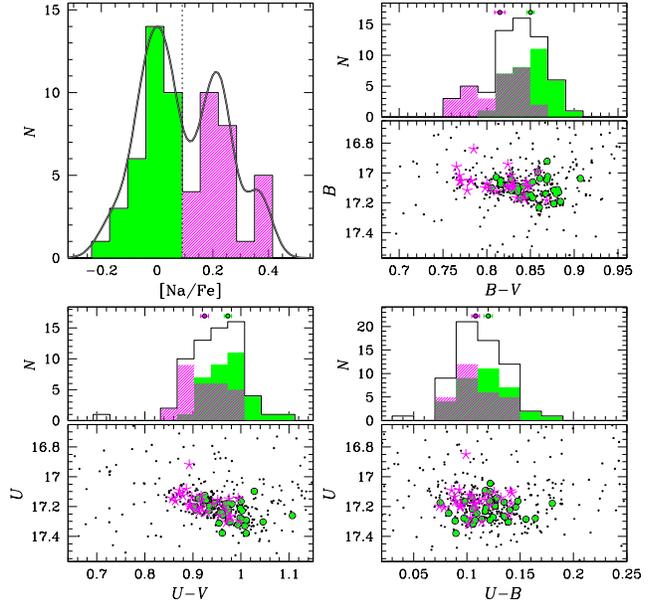}
      \caption{
\textit{Upper-left panel:} Histogram distribution of [Na/Fe] for RHB
stars. Green and magenta colors represent the two samples of Na-poor and
Na-rich RHB stars, respectively. A kernel density distribution has been
superimposed to the observed distribution.  
\textit{Upper-right and lower panels}: The $B$ versus $(B-V)$,
$U$ versus $(U-V)$, and $U$ versus $(U-B)$ CMD of NGC\,2808 around the
RHB.  
Na-rich and Na-poor RHB stars are represented with green dots and
magenta asterisks, respectively. We also show the color histogram distributions 
of the two Na-subsamples of the RHB stars, using the same color scheme. } 
         \label{CMDNa}
   \end{figure}

Figure~\ref{fig:NaDistr} reveals that the [Na/H] distribution of RHB stars is broad with some hints of bimodality. Similar conclusions can be inferred from the distribution of [Na/Fe] shown in the upper-left panel of Fig.~\ref{CMDNa}. These stars span an interval in [Na/Fe] of $\sim$0.6 dex, which is larger than that expected from observational errors alone ($\sim$0.07~dex, Tab.~\ref{gir_errors}). 

To further investigate the presence of multiple stellar populations along the RHB of NGC\,2808, we define two groups of stars on the basis of the [Na/Fe] distribution of Fig.~\ref{CMDNa}: {\it (i)} Na-rich RHB stars with [Na/Fe]$>$0.10~dex; and {\it (ii)} Na-poor RHB stars with [Na/Fe]$<$0.10~dex. These two groups of stars have been colored in magenta and green, respectively. These same color codes are used consistently hereafter. In the other panels of Fig.~\ref{CMDNa} we show the $B$ versus $(B-V)$ (upper-right panel), $U$ versus $(U-V)$ (lower-left panel), and $U$ versus $(U-B)$ (lower-right panel) CMDs in the RHB region. The position of spectroscopic targets in the CMD suggests that the Na-rich RHB stars have on average bluer colors with respect to the Na-poor stars. This behaviour is seen in the $(B-V)$, $(U-V)$ and $(U-B)$ colors. The mean differences (Na-rich$-$Na-poor) observed in various colors are: $\Delta(B-V)$=$-$0.035$\pm$0.007, $\Delta(U-V)$=$-$0.048$\pm$0.011, and $\Delta(U-B)$=$-$0.012$\pm$0.005.

Considering the working hypothesis of two groups of stars separated at [Na/Fe]=0.10~dex, the average sodium abundances for these groups are: [Na/Fe]=0.24$\pm$0.01~dex (rms=0.08), and [Na/Fe]=$-$0.02$\pm$0.01~dex (rms=0.07). This difference does not originate in our treatment of NLTE effects, as these corrections are comparable for the two groups. NLTE corrections for stars at higher Na are larger than those for low Na stars by only 0.02~dex. Since the NLTE corrections are fairly line-strength dependent, the small difference with respect to Na reflects that temperature and abundance influence line strength in opposite directions, partially cancelling with each other. In addition to this, the fact that Na-poor stars have, on average, redder colors than Na-rich ones further supports the idea that the RHB is not consistent with a simple stellar population. We note that systematics in the adopted \teff\ values are unlikely to produce the observed difference in [Na/Fe]. To cancel the observed variation in [Na/Fe] we would need \teff\ values for the Na-rich stars colder by $\sim$350~K (see Tab.~\ref{gir_errors}). This requires these stars to be redder by $\sim$0.15 in $(B-V)$, several times (seven) the observational error. As shown in Fig.~\ref{CMDNa}, such a large systematic in the $(B-V)$ should also affect the other investigated colors, which we do not observe.

\begin{figure}
   \centering
   \includegraphics[width=7cm]{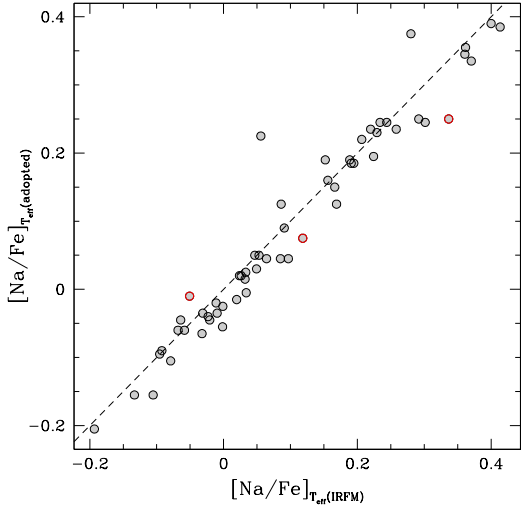}
   \caption{Comparison of the Na abundances relative to iron derived using our adopted atmospheric parameters and those derived using \teff\ obtained from the IRFM. Both sets of Na abundances are corrected for NLTE effects. The dashed line represents perfect agreement. Stars represented with red open circles have the largest uncertainties in the \teff\ derived from the IRFM ($\gtrsim$100~K).}   
   \label{fig:irfm_na}
   \end{figure}

As a final test, we have re-determined the Fe and Na abundances using temperatures derived from the IRFM and \logg\ consistent with these alternative \teff. In Fig.~\ref{fig:irfm_na} we compare the NLTE [Na/Fe] abundances derived using the two different sets of atmospheric parameters, e.g.\ the adopted ([Na/Fe]$_{\rm {\small \teff(adopted)}}$) and the IRFM ones ([Na/Fe]$_{\rm {\small \teff(IRFM)}}$). There is no significant systematic in the derived [Na/Fe], with a mean difference [Na/Fe]$_{\rm {\small \teff(IRFM)}}-$[Na/Fe]$_{\rm {\small \teff(adopted)}}=+0.01\pm0.01$~dex (rms=0.04), consistent with zero. More importantly, there is no evidence for systematics that affect different ranges in the observed [Na/Fe] distribution in different ways.

\begin{figure*}
   \centering
   \includegraphics[width=15.5 cm]{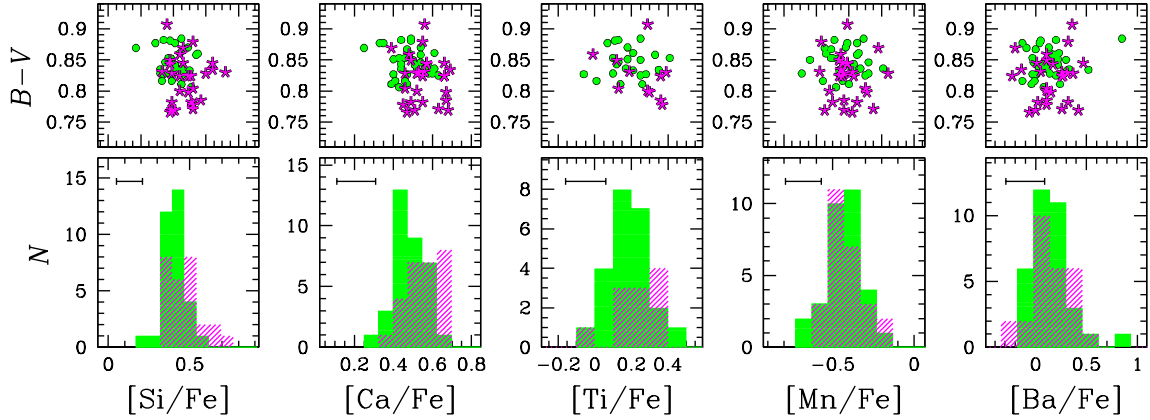}
      \caption{{\it Lower panels}: Distribution of the derived chemical
        abundances for the Na-rich (magenta histograms) and Na-poor RHB
        stars (green histograms) analysed with GIRAFFE. 
        {\it Upper panels}: $(B-V)$ as a function of  
        the chemical abundance. Colors and symbols are as in Fig.~\ref{CMDNa}.}
        \label{histo}
   \end{figure*}

We conclude that the RHB stars Na distribution is not consistent with a single stellar population of stars homogeneous in Na. Although there are hints of a bimodal distribution in Na in the RHB, present uncertainties do not allow us to determine whether or not the observed distribution translates into a continuous or discrete star-formation history. Apart from Na, the chemical composition of the two stellar populations along the RHB is uniform within observational errors. When we consider the groups as defined in the upper-left panel of Fig.~\ref{CMDNa}, there is no evidence of separate distributions in other chemical abundances. This is shown in Fig.~\ref{histo} where we plot the histograms of the silicon, calcium, titanium, manganese, and barium abundance distribution for Na-rich and Na-poor RHB stars.  

\begin{figure}
   \centering
   \includegraphics[width=5.8cm]{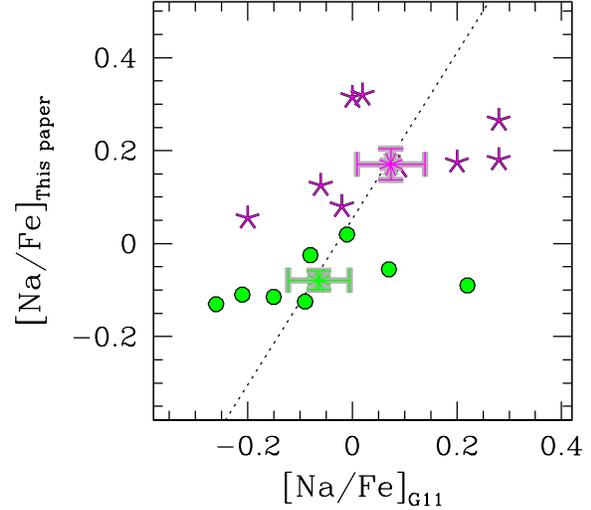}
      \caption{[Na/Fe] derived in this analysis compared to that from G11 for the seventeen stars in common. Colors and symbols are as in Fig.~\ref{CMDNa}. Super-imposed on the data are the average values for Na-rich and Na-poor stars, and the associated errors. The dotted-line represents the least-squares linear fit.  }
         \label{NaCompare}
   \end{figure}

Seventeen RHB targets analysed in this paper are in common with the sample of G11. In Fig.~\ref{NaCompare} we compare the [Na/Fe] inferred in this paper and those of G11 for the same stars. The stars have been divided into Na-poor and Na-rich subsamples according to our criterion; eight are Na-poor and nine Na-rich. The least squares fit with the data is shown as a dotted line. Both analyses exhibit similar dispersion, e.g.\ rms=0.16 (this paper), and rms=0.17 (G11), and a mean difference of $\Delta$([Na/Fe]$_{\rm this~paper-G11}$)=0.13$\pm$0.04 (rms=0.16). Although there is a positive correlation between our Na values and those from G11, in the G11 data, the stars we assign to the Na-rich and Na-poor group have, on average, similar [Na/Fe] (within $\sim$1~$\sigma$) with a large overlap. We do not know the reason for this difference, and several effects may contribute. 

\begin{figure}
   \centering
   \includegraphics[width=8.7cm]{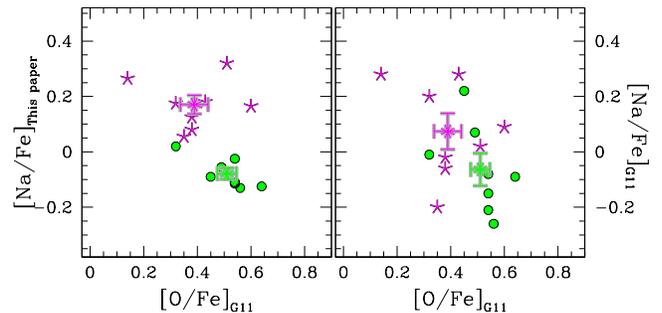}
      \caption{Sodium as a function of oxygen abundance 
        for seventeen stars in common with G11. Colors and symbols are as in Fig.~\ref{CMDNa}.
        The [Na/Fe] determinations shown in the left and right 
        panel are from this work and G11, respectively. Super-imposed on the data are the average values for Na-rich and Na-poor stars, and the associated errors.}
         \label{NaO}
   \end{figure}

To further strengthen our results, we examine the oxygen abundances for these common stars available from G11. Figure~\ref{NaO} shows [Na/Fe] as a function of [O/Fe] for these stars, using sodium abundance determinations from this paper (left panel) and from G11 (right panel). Na-rich stars are, on average, depleted in oxygen by 0.12$\pm$0.06 dex with respect to Na-poor stars. This difference, which is marginally significant (at the 2~$\sigma$ level), suggests that the RHB sample may have a mild Na-O anticorrelation, providing further evidence that it is not consistent with a simple stellar population.

%
%

%

\section{Helium}
\label{heBHB}

\begin{figure*}
   \centering
   \includegraphics[width=8.5 cm]{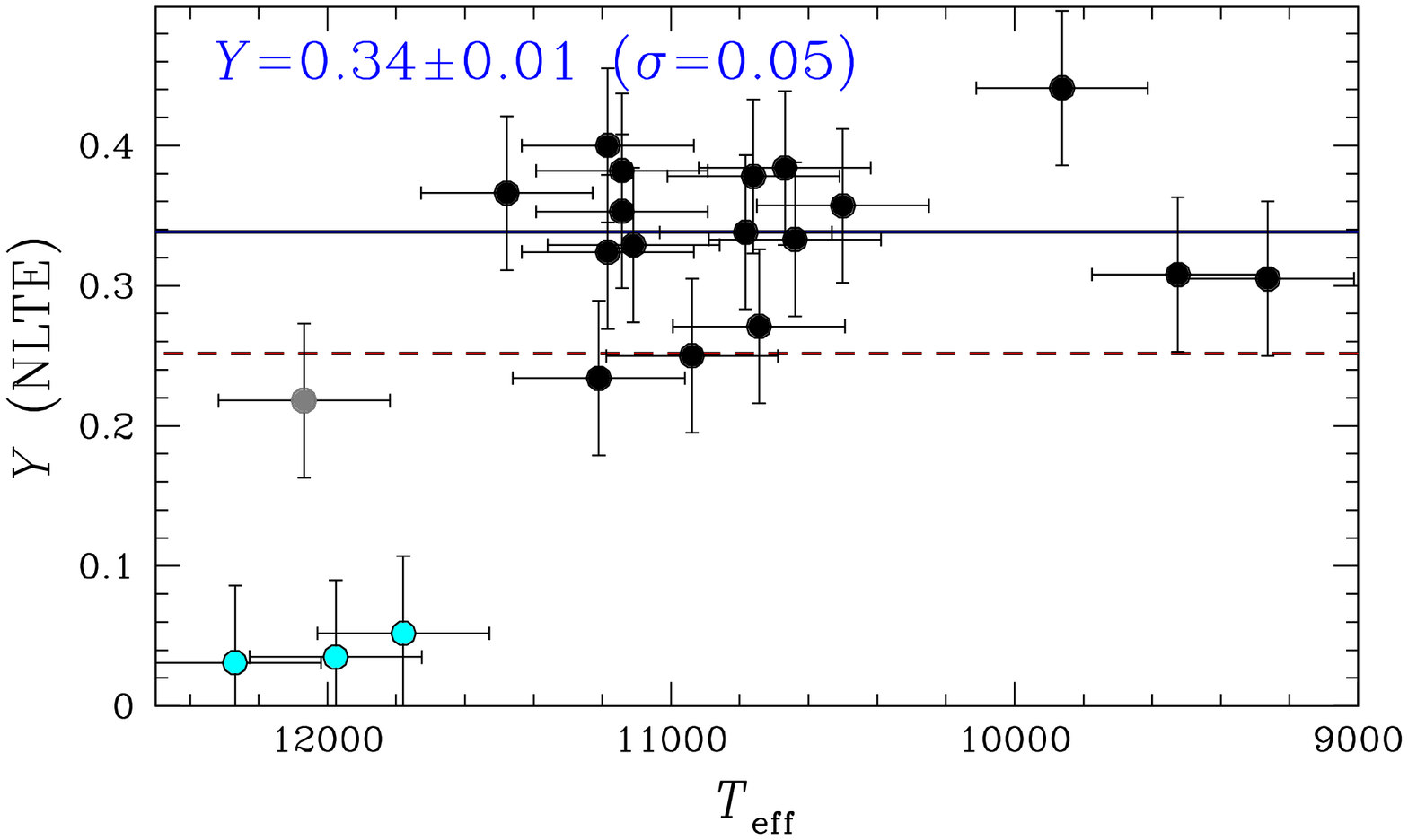}
   \includegraphics[width=6.0 cm]{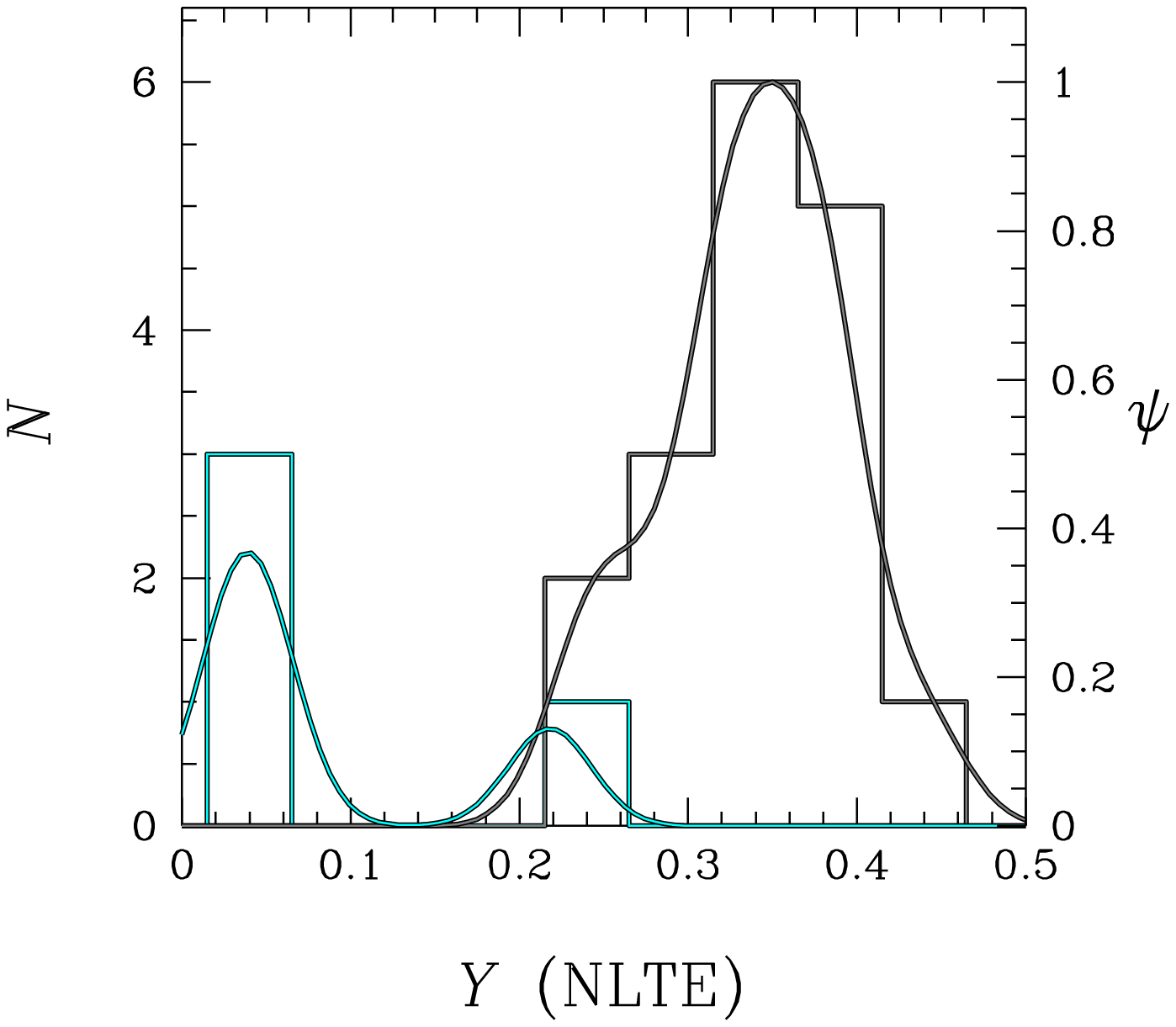}
      \caption{{\it Left panel}: 
        Helium mass fraction values, corrected for NLTE effects, as a function of
        effective temperature. The red-dashed line corresponds to
        primordial $Y$, while the blue solid line is the averaged value for
        the analysed blue HB stars, excluding those hotter than
        the Grundhal jump. Symbols are as in Fig.~\ref{he}. 
        {\it Right panel}: Histogram and kernel density distribution 
        of the $Y$ mass fractions, for stars colder (black) and warmer (cyan)
        than GJ.}
        \label{HePRZ}
   \end{figure*}

Spectral synthesis of the He lines, as explained in Sect.~\ref{bhbpar}, returns the He mass fractions corrected for NLTE effects listed in Tab.~\ref{tab:bhb_he}. Figure~\ref{HePRZ} shows NLTE values for $Y$ as a function of the effective temperature. Most stars with \teff$<$11500~K are enhanced in He with respect to the primordial value. Stars hotter than 11500\,K, for which we assume solar metallicity due to radiative levitation of metals, have $Y$\,$\lesssim$0.05. We attribute this drop in the $Y$ value to helium sedimentation that occurs at these temperatures. We set the primordial $Y$ level (dashed red line) to 0.2516, based on Izotov et al. (2007)\footnote{Izotov et al.\ (2007) report a helium content of $Y$=0.2472$\pm$0.0012 or $Y$=0.2516$\pm$0.0011, depending on the method used.}. The average He abundance for stars redder than the Grundahl jump is $Y$=0.34$\pm$0.01 (rms=0.05).

To estimate the internal error associated with the individual measurements of helium contents we re-compute the abundance determinations for a set of BHB stars that span the observed range in temperature, varying the atmospheric parameters by their expected errors (as explained in Sect.~\ref{atm_bhb}). The sensitivities of $Y$ to variations in stellar parameters, plus the errors introduced by uncertainties in the continuum placement, for these re-analysed stars are listed in Tab.~\ref{tab:He_errors}. By summing in quadrature the various contributions, we estimate the total internal uncertainty in $Y$ as 0.05-0.06, dominated by errors from temperature. Errors in microturbolence and metallicity are negligible, while gravities only marginally contribute to the abundance uncertainty.

Three stars in the BHB sample are consistent with primordial He. However, due to the relatively large internal uncertainty associated with the individual helium measurements, their $Y$ values are still consistent (within 2~$\sigma$ level) with the mean enhanced He derived for the whole sample. Hence, our data do not allow us to establish if these stars truly belong to a population with {\it normal} helium content. As a check, we determined \teff\  and \logg\ by using the photometric catalogs obtained from ACS/{\it HST} (Anderson et al.\ 2008), coupled with theoretical models, similarly to what was done by using WFI photometry (see Sect.~\ref{atm_bhb}). Photometry from space is more accurate but is available for a limited region of $\sim$3\arcmin$\times$3\arcmin, which includes only six of the BHB stars in our sample. Temperatures and gravities obtained with this photometry are listed in Tab.~\ref{tab:bhb_he} (last two columns). Atmospheric parameters from the two photometries are in agreement within expected errors, except for one star (\#16837) for which \teff\ from {\it HST} photometry is significantly lower. Note that, for this star, the \teff\ from {\it HST} agrees better with the EW of the He line, which lies among the values obtained for colder stars. Adopting a lower \teff\ would significantly increase the He abundance of this star; the $Y$ of the others would only marginally change.

Despite the relatively large internal uncertainty associated with the individual helium measurements, the statistic of our analysed sample provides some constraints on the $Y$ content of the analysed BHB stars. Indeed, the uncertainty associated with the average $Y$ (of $\pm$0.01) is relatively small, suggesting that BHB stars in NGC\,2808, in the portion of the HB studied here, are on average He enhanced.

To examine the reliability of this result, we investigate how sensitive it is to systematics that may affect the atmospheric parameters. As discussed in Sect.~\ref{atm_bhb}, from the theoretical side, we expect that some stars have a \logg\ lower than that adopted here. These differences in \logg, however, are expected to be at most $\sim$0.15~dex, and would lower the $Y$ measurements by no more than $\sim$0.02 (see Tab.~\ref{tab:He_errors}). Note that the $Y$ used to derive the atmospheric parameters is 0.02 lower than the mean value derived here. However, the resulting difference in \logg\ is relatively small and has a negligible impact on the $Y$ measurements, given our uncertainties. These arguments suggest that, even when we consider systematics in \logg, the average $Y$ obtained for the analysed BHB stars is still consistent with being higher than the primordial value.  

In comparison to \logg, systematics in \teff\ will cause larger variations in the derived $Y$ abundances. While we do not expect that the assumption on $Y$ systematically changes \teff\ values (see Sect.~\ref{atm_bhb}), there could be a systematic of $\lesssim$200~K with respect to the \teff\ scale adopted for the RHB, that is colder. The values listed in Tab.~\ref{tab:He_errors} (Systematic), suggest that this systematic uncertainty would increase $Y$ by $\sim$0.04. 

In summary, we conclude that our derived mean $Y$ for the analysed BHB stars colder than $\sim$11500~K is $Y$=0.34$\pm$0.01$\pm$0.05 (internal plus systematic uncertainty), higher than the primordial value. This result is robust within both our estimated internal and systematic errors. To shift the mean $Y$ value to a primordial value of $Y$=0.25 requires either the \teff-scale to be underestimated by $\sim$400~K or the \logg-scale to be overestimated by $\sim$0.60~dex. This result provides a direct confirmation of the predictions by D'Antona et al.\ (2004, 2005) that stellar populations enhanced in helium are present in the BHB of NGC\,2808 (see also Piotto et al.\ 2007; Dalessandro et al.\ 2011; Milone et al.\ 2012d). As mentioned above, our observational errors do not allow us to establish if the analysed sample has a uniform content of helium.

As discussed in Sect.~\ref{giraffe_bhb}, four stars show a significant drop in the EW of the helium line as a function of \teff, due to helium sedimentation. The helium content of these stars is sensibly low, ranging from $Y \sim$0.03 to $Y \sim$0.21. 

\section{Discussion}
\label{discussion}
  
The presence of multiple stellar groups in NGC\,2808 has been inferred along all the evolutionary stages of the CMD, both spectroscopically and photometrically, implying the co-existence of at least three populations of stars. The observed triple MS (Piotto et al.\ 2007) provides indirect information on the He content, implying the presence of:
\begin{itemize}
\item{A population defining the red MS characterised by primordial He, hereafter first-generation;}
\item{An intermediate He population defining the middle MS associated with $Y$$\sim$0.32, hereafter intermediate second-generation;}
\item{A highly He-enhanced population distributed on the blue MS with $Y$$\sim$0.38, hereafter extreme second-generation.}
\end{itemize}

The different He content of the three NGC\,2808 populations was first predicted by D'Antona et al. (2005) from their analysis of HB and MS stars. According to this work, stars with different He content populate different HB segments. The distribution of stars along the HB in this cluster shows three significant gaps and distinct segments, RHB, EBT1, EBT2, and EBT3 (Sosin et al.\ 1997; B00). Consistent with the predictions of D'Antona et al. (2005), stars occupying the RHB of NGC\,2808 are associated with first generation stars, and stars in the EBT1 and EBT2$+$EBT3 HB segments are representative of the intermediate second generation and the extreme second generation, respectively. The Na-O anti correlation is also consistent with three groups in oxygen (Carretta et al. 2007), corresponding to three RGBs (Monelli et al.\ 2013) and providing indirect evidence of He enrichment in NGC\,2808. Note that an alternative scenario for the hottest stars of the HB, the EBT3 stars, was proposed by Moehler et al.\ (2011). They found that the hot end of the HB in $\omega$~Centauri (the so-called blue-hook) is not solely populated by He-rich stars and suggest that stars in this extremely hot HB segment may be formed as a consequence of independent evolutionary channels (like hot He flashers; see also Cassisi et al.\ 2009). 

To explain the three stellar populations in NGC\,2808, D'Ercole et al. (2008, 2010) propose that a highly He-rich and O-poor extreme second generation of stars, corresponding to the blue MS, formed directly from the pure ejecta of a first generation of AGBs. This happened after the proto-cluster was cleared from residual gas by SN feedback or by dark-remnant accretion (Krause et al.\ 2012). After this gas expulsion epoch, a cooling flow sets in and brings to the core the low velocity ejecta of AGB stars. Later on, the massive AGB ejecta is mixed with pristine gas (re-accreted in the cluster core), giving origin to the "intermediate'' SG stars, with milder O depletion and He enhancement.

{\it According to this picture, our results suggest that the evolution of this cluster could be even more complex than a three population scenario.} We find that the RHB is consistent with stars having different Na and little difference in O. If we assume that the RHB stars are the progeny of the red MS stars, our results imply that the red MS is not composed of a single stellar population, but includes stars with some spread in the light element abundances. Interestingly, by analysing high precision {\it HST} data for the MS of NGC\,2808, Milone et al.\ (2012d) noted that the red MS is broader than expected from observational errors and/or binaries, and its spread in color at various magnitudes is larger than that associated with the intermediate and blue MS. This broadening in the $(F475W-F814W)$ {\it ACS/WFC} filters (Piotto et al.\ 2007; Milone et al.\ 2012d) is consistent with a small spread in $Y$ among red MS stars. The mean higher luminosity of Na-rich RHB stars we observe (see Fig.~\ref{CMDNa}) is consistent with a small difference (spread) in the He content in the RHB of NGC\,2808, e.g.\ He rich and Na-rich RHB stars are brighter.  

This observational scenario qualitatively agrees with results on the RHB of the metal-rich GC 47~Tucanae. The HB stars of this cluster ([Fe/H]=$-$0.72) are distributed on the red side of the instability strip, consistent with a single segment. However, CN-strong stars observed in the GC, which are expected to be enriched in elements produced by high temperature H-burning (like Na), are slightly more luminous and consistent with having higher He. The analysis of the MS for this cluster suggests that the maximum variation in $Y$ for different stellar populations is small, at most $\sim$0.02 (Milone et al.\ 2012a).

Our findings on NGC\,2808 present a difficult observational puzzle. The RHB, expected to be the progeny of the red MS, hosts stars that underwent an enrichment in Na, along with first-generation Na-poor RHB stars. Given that no metallicity variations have been observed for NGC\,2808, the spread on the red MS, observed in photometric filters not seriously affected by molecular bands, may be caused by a small $Y$ spread (Sbordone et al. 2011; Cassisi et al. 2013). This prompts the question of when these mildly polluted stars formed in the evolution of the cluster. Following the D'Ercole et al. scenario, we suppose that they are the latest stars to have been formed from highly diluted material, such that their abundances in light elements and $Y$ approach the primordial values of the first generation. Hence, these stars could have formed after the intermediate second generation stars, which show evidence for a higher degree of AGB pollution. If this prediction is correct, the abundance pattern of light elements of these stars is dominated by dilution with pristine gas, and the red MS and the RHB contain the first and last stars formed in the cluster. 

A bi-modal or broad distribution in the chemical abundances of light elements, such as Na, among RHB stars would provide support to this scenario. Stars belonging to the first generation and the most-recently formed stellar population should record the abundances of the interstellar medium from which they were born, and thus, are expected to be different from one another. Some hints of possible bi-modality in the Na content of our sample of RHB stars are present, but the relatively poor statistics and internal uncertainties of our data prevent us from making definitive conclusions on the presence of two discrete Na groups or an intrinsically broad distribution. In the latter case, the star formation in the final stages of the cluster evolution may have occurred continuously, not in discrete bursts.

As confirmed by the higher Na content, second generation(s) stars, which formed from material enriched in the high-temperature H-burning products, populate the BHB. This is analogous to observations of M\,4 (Marino et al.\ 2011a). The He abundances of the hottest BHB stars in our sample suggest that they are enhanced in the He mass fraction $Y$ by $\sim$0.09 with respect to the primordial value. This is a strong direct spectroscopic measurement of an enhanced He content for second generation stars in a GC. Our analysis takes NLTE effects into account, which is a imperative to determine an accurate abundance of He in BHB.

The mean $Y$ found here agrees with the value predicted from models for the middle MS. The extreme population of NGC\,2808 (corresponding to the blue MS) is predicted to have even larger He enhancement, up to $Y \sim$0.40, and is expected to populate the EBT3 segment on the HB (D'Antona et al.\ 2005; Dalessandro et al.\ 2011). The He abundances determined in this work are sensitive to systematics in our \teff\ and \logg\ scale. We pay close attention to the calibration of our atmospheric parameters, which seem to be correct within $\sim$200~K in temperature and $\sim$0.20~dex in gravity, and we remind the reader that the He enhancement of BHB stars in NGC\,2808 is valid if our \teff\ and \logg\ scales are correct within 400~K and 0.60~dex, respectively.

\section{Conclusions}
\label{conclusions}

We have reported chemical abundance analysis for HB stars in the GC NGC\,2808. From our study we conclude that: 
\begin{itemize} 
\item{the presence of second generation(s) helium-enhanced stars is strongly supported by the mean He abundance derived for BHB stars;}
\item{in NGC\,2808, He enriched stars occupy the BHB, corroborating the idea that He is a fundamental parameter for the HB morphology; }
\item{BHB stars have higher Na than RHB stars; }
\item{RHB stars show internal variations in Na, with hints of a possible bi-modality, demonstrating that not even this sub-population is consistent with a single stellar population.}
\end{itemize} 

Our findings on NGC\,2808 RHB stars fit within the scenario of multiple stellar populations in GCs, where Na-rich/O-poor stars are polluted by processed gas from a previous generation of stars and are representative of the He enhanced population(s). 

That the RHB is not consistent with a simple stellar population suggests that NGC\,2808 could have experienced a very complex star-formation history with more than three stellar generations. This supports photometric studies that show both red and blue MS exhibiting internal $m_{\rm F475W}-m_{\rm F814W}$ color spread, as expected if their stars are not chemically homogeneous (Piotto et al.\ 2007, Milone et al.\ 2012d).  

Stars with normal O and Na values are representative of the unpolluted He normal population. Variation in helium is expected to manifest as peculiar observed features of the CMD, such as multiple main sequences and HB morphology.  Empirical detection of He enhancement in GC stars constitutes the essential confirmation of this scenario. Pasquini et al. (2011) found spectroscopic evidence for a possibly high difference in He for two stars on the RGB of NGC\,2808. A large difference in He has also been inferred for two RGBs in the peculiar cluster $\omega$~Centauri (Dupree \& Avrett 2013). The present paper provides the first direct spectroscopic measurement of highly ($Y \sim$0.34) He-enhanced stars in the BHB of a GC, confirming that He-enrichment occurred in GCs, and that He content guides the distribution of stars along the HB.  

\section*{acknowledgments}

\small
We are grateful to Y. Momany for providing his photometric catalog.
We thank the anonymous referee for his/her helpful comments that improved the quality of the paper.
APM and HJ acknowledge the financial support from the Australian Research 
Council through Discovery Project grant DP120100475.
APM, GP, SC are founded by the Ministry of Science and
Technology of the Kingdom of Spain (grant AYA 2010-16717). 
APM is also founded by the Instituto de Astrof{\'{\i}}sica de Canarias (grant P3-94).
MC 
and MZ are
supported by Proyecto Fondecyt Regular \#1110326 and \#1110303; 
the BASAL Center for Astrophysics and Associated
Technologies (PFB-06); the FONDAP Center for Astrophysics 15010003; 
Proyecto Anillo ACT-86 and
by the Chilean Ministry for the Economy, Development, and
Tourism's Programa Iniciativa Cient\'{i}fica Milenio through grant P07-021-F, 
awarded to The Milky Way Millennium Nucleus;
AARV acknowledges additional support from CAPES grant PNPD/2011-Institucional.
Support for R.A. is provided by Proyecto GEMINI-CONICYT 32100022 and via
a Postdoctoral Fellowship by the School of Engineering at Pontificia
Universidad Cat{\'{o}}lica de Chile.


\onecolumn

\begin{table*}
\caption{Basic data for the analysed UVES and GIRAFFE stars. Identification numbers and photometric data are from Momany et al. (2004).}
\begin{tabular}{rccccccclcr}
\hline\hline
\label{phot_data_tab}
 ID    &  $\alpha$(2000) & $\delta$(2000)    &   $V$ & $(B-V)_{\rm {ori}}$  & $(B-V)_{\rm {corr}}$&RV [\kmsec]& ID(2MASS) & type  \\
\hline
\\
\multicolumn{9}{c}{UVES}\\
\\
21854  &   09:12:04.408    & $-$64:50:47.662 & 16.256 & 0.829 & 0.843 & 107.1  & --               & RHB \\ 
18106  &   09:12:02.698    & $-$64:52:31.346 & 16.249 & 0.815 & 0.814 & 109.4  & --               & RHB  \\
20577  &   09:11:55.598    & $-$64:51:21.500 & 16.100 & 0.781 & 0.789 & 102.4  & --               & RHB  \\
  716  &   09:12:03.965    & $-$64:49:26.226 & 16.705 & 0.317 & 0.298 & 106.0  & --               & BHB  \\
19229  &   09:12:23.239    & $-$64:51:59.650 & 16.290 & 0.212 & 0.232 & 110.5  & 09122323-6451595 & BHB  \\
  509  &   09:12:18.219    & $-$64:49:37.130 & 16.768 & 0.189 & 0.180 & 116.8  & 09121814-6449371 & BHB  \\
11715  &   09:11:58.235    & $-$64:57:13.035 & 16.917 & 0.132 & 0.121 & 105.5  & 09115825-6457129 & BHB  \\
\\
\multicolumn{9}{c}{GIRAFFE}\\          
\\                             
2771     &   09:11:54.180    & $-$64:47:10.765 & 16.236 & 0.844 & 0.828 &   95.3 & --                &    RHB  \\
2648     &   09:11:44.296    & $-$64:47:23.509 & 16.296 & 0.845 & 0.827 &  100.5 & 09114429-6447236  &    RHB  \\
4036     &   09:11:39.084    & $-$64:49:51.110 & 16.290 & 0.925 & 0.869 &  112.8 & 09113906-6449509  &    RHB  \\
4491     &   09:11:40.148    & $-$64:49:04.720 & 16.356 & 0.808 & 0.769 &  111.9 & 09114013-6449046  &    RHB  \\
3093     &   09:12:09.296    & $-$64:46:32.395 & 16.289 & 0.845 & 0.835 &   94.4 & 09120929-6446324  &    RHB  \\
1317     &   09:12:02.282    & $-$64:48:55.849 & 16.334 & 0.872 & 0.849 &  107.8 & 09120228-6448558  &    RHB  \\
\hline        
\end{tabular}                                                                                          
\end{table*}

\begin{table*}
\caption{Atmospheric parameters from Fe lines, and \teff\ from photometry for                                            
 the RHB stars observed with UVES.}
\label{uves_atm}
\begin{tabular}{lcccccc}                       
\hline\hline
ID&\teff &\logg&\vmicro & \teff\ (Alonso) \\
\multicolumn{1}{c}{}&\multicolumn{3}{c}{Fe lines}
&\multicolumn{1}{c}{photometry}\\
\hline

20577 & 5390 & 2.26 & 1.66  & 5501\\
18106 & 5480 & 2.55 & 1.48  & 5422\\
21854 & 5350 & 2.20 & 1.73  & 5336\\

\hline
\end{tabular}
\end{table*}


\begin{table*}
\scriptsize
\caption{Atmospheric parameters and chemical abundances for the RHB stars and colder BHB stars observed with GIRAFFE. For Na and Mn we list the line-to-line scatter.}
\label{gir_par_tab}
\begin{tabular}{cccccccrcccrccrc}
\hline\hline
 ID      &  \teff   & \logg &  \vmicro & ${\rm [Fe/H]}${\sc i} & ${\rm [Fe/H]}${\sc ii} & ${\rm [Na/Fe]}_{\small {\rm LTE}}$ & ${\rm [Na/Fe]}_{\small {\rm NLTE}}$ & err$_{\small {\rm [Na/Fe]}}$ & ${\rm [Si/Fe]}$ & ${\rm [Ca/Fe]}$ & ${\rm [Ti/Fe]}$ & ${\rm [Mn/Fe]}$ & err$_{\small {\rm [Mn/Fe]}}$ & ${\rm [Ba/Fe]}${\sc ii}  & \teff\ ${\rm {\small (IRFM)}}$ \\ 
\hline
2771     &   5381   & 2.34  &    1.54  & $-$1.32    & $-$1.24  &  0.49   &   0.04 & 0.04  &  0.32 &  0.60 &   0.10   & $-$0.28 & --  &  0.22  &   --   \\
2648     &   5384   & 2.36  &    1.54  & $-$1.07    & $-$0.92  &  0.39   &   0.01 & 0.04  &  0.34 &  0.40 &$-$0.06   & $-$0.37 & --  &  0.15  &  5452   \\ 
4036     &   5262   & 2.30  &    1.55  & $-$1.22    & $-$1.14  &  0.21   &$-$0.20 & 0.15  &  0.17 &  0.25 & --       & $-$0.55 & --  & $-$0.01&  5384   \\ 
4491     &   5563   & 2.40  &    1.53  & $-$1.11    & $-$0.98  &  0.76   &   0.39 & 0.10  &  0.41 &  0.51 & --       & $-$0.55 & --  &  0.55  &  5614   \\ 
3093     &   5360   & 2.35  &    1.54  & $-$1.19    & $-$1.17  &  0.34   &$-$0.09 & 0.00  &  0.32 &  0.44 & --       & $-$0.45 & 0.09&  0.35  &  5448   \\ 
1317     &   5319   & 2.34  &    1.54  & $-$1.25    & $-$1.13  &  0.57   &   0.23 & 0.05  &  0.38 &  0.49 &   0.12   & $-$0.46 & --  &  0.22  &  5479   \\ 
\hline                                                                                                  
\end{tabular}                                                                                         
\end{table*}


\begin{table*}
\scriptsize
\caption{Atmospheric parameters and individual abundances, with the associated line-to-line scatter, for the RHB stars observed with UVES.}
\begin{tabular}{cccccrrrrrrr}
\hline
\hline
ID & \teff & \logg & \vmicro & [Fe/H] & [Fe/H]{\sc i}   &  [Fe/H]{\sc ii} & [Na/Fe]$\rm {_{orig}}$ & [Na/Fe]$\rm {_{corr}}$ & [Mg/Fe]  & [SiFe] & [Ca/Fe] \\
\hline
20577 & 5390 & 2.26 & 1.66 & $-$1.21 & $-$1.24$\pm$0.03 & $-$1.17$\pm$0.12 & 0.23$\pm$0.19 & $-$0.13$\pm$0.08 & 0.23$\pm$0.10 & 0.24$\pm$0.13 &0.20$\pm$0.06\\
18106 & 5480 & 2.55 & 1.48 & $-$1.14 & $-$1.18$\pm$0.03 & $-$1.10$\pm$0.07 & 0.35$\pm$0.16 &       0.03$\pm$0.04 & 0.29$\pm$0.11 & 0.18$\pm$0.10 &0.15$\pm$0.06\\
21854 & 5350 & 2.20 & 1.73 & $-$1.17 & $-$1.21$\pm$0.03 & $-$1.12$\pm$0.08 & 0.11$\pm$0.08 & $-$0.27$\pm$0.13 & 0.31$\pm$0.08 & 0.12$\pm$0.12 &0.24$\pm$0.06 \\
       &         &         &          &               &                                  &                                 &                           &                              &                              &                        &                           \\ 
\end{tabular}
\begin{tabular}{rrrrrrrrrr}

[Sc/Fe]{\sc ii} & [Ti/Fe]{\sc i} &  [Ti/Fe]{\sc ii} & [Cr/Fe]{\sc i} & [Cr/Fe]{\sc ii}  & [Ni/Fe]& [Zn/Fe] & [Y/Fe]{\sc ii} & [Ba/Fe]{\sc ii} & [Nd/Fe]{\sc ii}  \\
\hline
$-$0.01$\pm$0.05 &  0.34$\pm$0.01 & 0.20$\pm$0.06 & $-$0.19$\pm$0.04 &       0.12$\pm$0.00 & $-$0.07$\pm$0.07 & $-$0.18$\pm$0.00 & $-$0.20$\pm$0.04 & $-$0.14$\pm$0.00 &  0.15$\pm$0.00 \\
     0.17$\pm$0.08 &  0.39$\pm$0.11 & 0.22$\pm$0.15 &$-$0.13$\pm$0.06  & $-$0.18$\pm$0.00 & $-$0.04$\pm$0.12 &       0.16$\pm$0.00 & $-$0.11$\pm$0.04 &  0.12$\pm$0.00      &  0.01$\pm$0.00             \\
     0.07$\pm$0.08 &  0.25$\pm$0.08 & 0.13$\pm$0.10 & $-$0.11$\pm$0.11 & $-$0.04$\pm$0.00 & $-$0.03$\pm$0.05 & $-$0.02$\pm$0.00 & $-$0.22$\pm$0.07 & $-$0.04$\pm$0.00 & $-$0.03$\pm$0.00             \\
\hline
\label{uves_abb}
\end{tabular}
\end{table*}


\begin{table*}
\caption{Sensitivity of derived GIRAFFE RHB abundances to the                                                         
  atmospheric parameters and EWs. We reported the total error due to                                                  
  the atmospheric parameters and the EW measurement, 
  the squared sum of these
  contributions ($\sigma_{\rm total}$), and the observed dispersion                                                  
  ($\sigma_{\rm obs}$) for each element. \label{gir_errors}}
\begin{tabular}{lccccccc}
\hline\hline
      &$\Delta$\teff &$\Delta$\logg&$\Delta$\vmicro&$\Delta$[A/H]& $\sigma_{\rm EW}$&$\sigma_{\rm total}$&$\sigma_{\rm obs}$\\  
      & $\pm$50~K    & $\pm$0.10   & $\pm$0.15~\kmsec & 0.07~dex &$\pm$4m\AA       &                   &\\
\hline
{\rm [Na/Fe]}{\sc i}   & $\pm$0.03   &$\mp$0.03  &$\mp$0.01  &$\pm$0.01 & $\pm$0.05 & 0.07 & 0.15\\
{\rm [Si/Fe]}{\sc i}   & $\mp$0.01   &$\pm$0.00  &$\pm$0.01  &$\pm$0.00 & $\pm$0.08 & 0.08 & 0.09\\
{\rm [Ca/Fe]}{\sc i}   & $\pm$0.01   &$\mp$0.02  &$\mp$0.06  &$\mp$0.01 & $\pm$0.07 & 0.10 & 0.09\\
{\rm [Ti/Fe]}{\sc i}   & $\pm$0.02   &$\pm$0.00  &$\pm$0.01  &$\pm$0.00 & $\pm$0.11 & 0.11 & 0.12\\
{\rm [Mn/Fe]}{\sc i}   & $\pm$0.01   &$\pm$0.00  &$\pm$0.02  &$\pm$0.00 & $\pm$0.11 & 0.11 & 0.10\\
{\rm [Fe/H]}{\sc i}    & $\pm$0.04   &$\pm$0.00  &$\mp$0.03  &$\pm$0.00 & $\pm$0.04 & 0.07 & 0.07\\
{\rm [Fe/H]}{\sc ii}   & $\pm$0.00   &$\pm$0.04  &$\mp$0.01  &$\pm$0.01 & $\pm$0.11 & 0.12 & 0.08\\
{\rm [Ba/Fe]}{\sc ii}  & $\pm$0.03   & $\mp$0.01 &$\mp$0.10  &$\pm$0.00 & $\pm$0.16 & 0.19 & 0.19\\
\hline
\end{tabular}
\end{table*}

      
\begin{table*}
\caption{Atmospheric parameters, chemical abundances and line-to-line scatter (when applicable) for the BHB UVES stars\label{BHB_TAB-UVES}}
\begin{tabular}{rrccccc}
\hline\hline
 ID      & \teff & \logg&\vmicro &[Fe/H]{\sc ii}      & [Na/Fe]$_{\rm NLTE}$ & [Mg/Fe]         \\\hline     
19229    & 9268  & 3.25 & 2.26   & $-$1.04$\pm$0.06     & 0.44                  & 0.12$\pm$0.05 \\         
  716    & 10144 & 3.51 & 2.07   & $-$1.23$\pm$0.10     & $<$0.85               & 0.21$\pm$0.08 \\         
  509    & 10387 & 3.59 & 2.02   & $-$1.15$\pm$0.10     & $<$0.86               & 0.16$\pm$0.04 \\         
11715    & 11036 & 3.74 & 1.90   & $-$1.07$\pm$0.11     & $<$0.80               & 0.39$\pm$0.07 \\         
\hline
\end{tabular}
\end{table*}


\begin{table*}
\caption{Atmospheric parameters, EWs for the He line, and He mass fractions 
  for the BHB stars observed with GIRAFFE.\label{tab:bhb_he}}
\begin{tabular}{ccccrrrrrrrrccc} 
\hline
\hline
 ID  &\teff &  \logg & \vmicro & EW & \vsini& $Y_{\rm NLTE}$          &   & \teff\ ($HST$)& \logg\ ($HST$)\\
     &  [K] & (cgs)  & \kmsec  & m\AA&\kmsec& mass fraction          &   & [K]           &  (cgs)        \\ 
\hline
17132  &   9263   &   3.27  &  2.26 &  18.7  &      0    &  0.305 & &  9312 & 3.27 \\ 
17826  &   9525   &   3.39  &  2.20 &  24.3  &     10    &  0.308 & &  9708 & 3.43 \\ 
10342  &   9862   &   3.51  &  2.13 &  60.2  &      0    &  0.441 & &       &      \\ 
 3030  &  10500   &   3.59  &  2.00 &  67.3  &      0    &  0.357 & &       &      \\ 
  231  &  10639   &   3.60  &  1.97 &  64.6  &      8    &  0.333 & & 10697 & 3.64 \\ 
17336  &  10669   &   3.68  &  1.97 &  76.2  &      0    &  0.384 & &       &      \\ 
11562  &  10744   &   3.66  &  1.95 &  61.2  &      0    &  0.271 & &       &      \\ 
23393  &  10760   &   3.68  &  1.95 &   84.7 &      0    &  0.378 & &       &      \\ 
14629  &  10783   &   3.67  &  1.95 &   67.4 &      0    &  0.338 & &       &      \\ 
16837  &  10939   &   3.63  &  1.92 &   62.3 &      0    &  0.250 & & 10003 & 3.49 \\ 
14968  &  11109   &   3.75  &  1.89 &   96.0 &     15    &  0.329 & &       &      \\ 
10883  &  11143   &   3.81  &  1.88 &  106.9 &      0    &  0.382 & &       &      \\ 
15081  &  11143   &   3.76  &  1.88 &   98.9 &      5    &  0.353 & &       &      \\ 
 1129  &  11184   &   3.81  &  1.87 &  103.8 &     10    &  0.400 & &       &      \\ 
23610  &  11184   &   3.77  &  1.87 &   93.5 &     10    &  0.324 & &       &      \\ 
20630  &  11210   &   3.78  &  1.87 &   62.6 &     10    &  0.234 & & 11189 & 3.76 \\ 
 1169  &  11478   &   3.85  &  1.82 &  124.2 &      0    &  0.366 & &       &      \\ 
\hline 
\multicolumn{10}{c}{Grundahl jump stars }\\
13305 & 11779  & 3.81 & 0.00  & 49.9  &  10  & 0.052  &  &       &      \\  
18640 & 11951  & 3.83 & 0.00  & 30.2  &   0  & 0.035  &  & 11604 & 3.84 \\ 
 4129 & 12068  & 3.96 & 0.00  & 122.3 &  10  & 0.218  &  &       &      \\ 
13132 & 12269  & 3.89 & 0.00  & 39.6  &  10  & 0.031  &  &       &      \\ 
\hline
\end{tabular}
\end{table*}


\begin{table*}
\caption{Sensitivity of NLTE helium abundances (mass fractions $Y$) to
  variations of atmospheric parameters and continuum setting, 
exemplarily for several BHB sample stars.\label{tab:He_errors}}
\begin{tabular}{lrrrrrrr}
\hline\hline
                               &    17132 &  10342    & 3030      &  14968   &  1169    &  13305    &  4129\\
Internal                       &          &           &           &          &          &           &          \\\hline
$\Delta \teff = +250$\,K       &  $-$0.055&  $-$0.055 &  $-$0.054 &$-$0.048  & $-$0.049 &  $-$0.013 &  $-$0.044 \\  
$\Delta \logg = +0.05$\,dex    &    +0.007&    +0.008 &    +0.008 &  +0.007  &   +0.009 &    +0.004 &    +0.011\\
$\Delta \vmicro = +1.00$\kmsec &   $-$0.002&    $-$0.001 & $-$0.002 &  $-$0.002  &   $-$0.002 &    +0.000 &    $-$0.002\\
$\Delta [$A/H$] = +0.07$\,dex  &    +0.000&    +0.000 &    +0.000 &  +0.000  &   +0.000 &    +0.000 &    +0.000\\
$\sigma_\mathrm{cont}$         &$\pm$0.015&$\pm$0.015 &$\pm$0.015 &$\pm$0.015&$\pm$0.015& $\pm$0.015& $\pm$0.015\\
$\sigma_\mathrm{total}$        &     0.057&     0.058 &     0.057 &    0.051 &    0.052 &     0.020 &     0.048\\
\hline
                               &          &           &           &          &          &           &          \\
Systematic                     &          &           &           &          &          &           &          \\\hline
$\Delta \teff = +200$\,K       &  $-$0.045&  $-$0.045 &  $-$0.043 &$-$0.039  & $-$0.039 &  $-$0.010 &  $-$0.036 \\  
$\Delta \logg = +0.15$\,dex    &    +0.020&    +0.022 &    +0.022 &  +0.019  &   +0.025 &    +0.012 &    +0.033\\
$\sigma_\mathrm{total}$        &     0.049&     0.050 &     0.048 &   0.043  &    0.046 &     0.017 &     0.049\\
\hline
\end{tabular}
\end{table*}

\end{document}